\documentclass[]{spie}  
\usepackage[dvips]{graphicx}

\title{X-ray Scattering from Random Rough Surfaces}

\author{Ping Zhao
\skiplinehalf
Harvard-Smithsonian Center for Astrophysics\\
60 Garden Street, Cambridge, MA 02138  U.S.A.}

\authorinfo{Further author information: Ping Zhao: E-mail:
zhao@cfa.harvard.edu}

\newcommand{\zs}{PZ\&LVS}
\newcommand{\pz}{PZ}
\usepackage{float}
\begin{document} 
\maketitle 

\begin{abstract} 

  This paper presents a new method to model X-ray scattering on random
  rough surfaces.  It combines the approaches we presented in two
  previous papers -- \zs\cite{zhao03} \& \pz\cite{zhao15}.  An
  actual rough surface is (incompletely) described by its Power
  Spectral Density (PSD).  For a given PSD, model surfaces with the
  same roughness as the actual surface are constructed by preserving
  the PSD amplitudes and assigning a random phase to each spectral
  component.  Rays representing the incident wave are reflected from
  the model surface and projected onto a flat plane, which is the
  first order approximation of the model surface, as outgoing rays and
  corrected for phase delays.  The projected outgoing rays are then
  corrected for wave densities and redistributed onto an uniform grid
  where the model surface is constructed.  The scattering is then
  calculated using the Fourier Transform of the resulting
  distribution.  This method provides the exact solutions for
  scattering in all directions, without small angle approximation.  It
  is generally applicable to any wave scatterings on random rough
  surfaces and is not limited to small scattering angles.  Examples
  are given for the Chandra X-ray Observatory optics.  This method is
  also useful for the future generation X-ray astronomy missions.

\end{abstract}


\section{INTRODUCTION}
\label{sect:intro}  

The study of wave scattering from rough surfaces goes back at least to
Rayleigh, in his 1877 classic -- {\it The Theory of
  Sound}\cite{rayleigh}, which led to the development of the Rayleigh
criterion (see Section \ref{sec:rrs}) for classifying the degree of
surface roughness.  Since then, the problem of scattering from random
rough surfaces has been investigated by many physicists and engineers,
and has been the subject of many books, including the classic -- {\it
  The Scattering of Electromagnetic Waves From Rough Surfaces} by
Beckmann and Spizzichino\cite{beckmann} and countless research
papers\cite{ogilvy}.  A good solution
for X-rays scattering in grazing incidence on random rough surfaces is
in high demand due to the emerging fields of X-ray optics in many
applications.  This problem is even more difficult due to the short
wavelength (comparing to the scale of the surface roughness) and the
small angle between the wave propagating direction and the surface.
Most approaches in the literature make the approximation that the
scattering angle is much smaller than the incident grazing angle.
Some of the treatments use the approximation that the surfaces are
sufficiently ``smooth'' so that a low order expansion in the surface
height errors is adequate, and consequently are limited in their
applications.  Most of those methods can not obtain the scattering
asymmetry around the direction of specular reflection (scattering
towards versus away from the surface).  These approximations are not
adequate for many of the applications involving X-ray mirrors.

In 2002, we presented a SPIE paper -- {\it A new method to model X-ray
  scattering from random rough surfaces} (\zs)\cite{zhao03} which
introduces a novel approach to the problem and gives the solution for
scattering in the incident plane.  In 2015, we presented a second SPIE
paper -- {\it Transverse X-ray scattering on random rough surfaces}
(\pz)\cite{zhao15} which gives the solution of scattering in the
direction perpendicular to the incident plane.  Based on the above two
SPIE papers, this paper provides a complete and exact solution for
X-ray scattering from random rough surfaces in all directions.  This
new method is generally applicable and provides the exact solution to
all wave scattering problems on random rough surfaces.

In previous methods, scattered wave was usually treated separately as
coherently reflected wave in the specular direction and incoherently
scattered (or diffused) wave in other directions (see
e.g. Simonsen\cite{simonsen}).  However, for any given wave, there is
no clear distinction between ``smooth'' and ``rough'' surfaces (see
Section~\ref{sec:rrs}), therefore it is impossible to separate
scattered wave into coherently reflected and incoherently scattered
waves.

The novelty of our new method is that it treats the reflected wave and
scattered wave together (e.g. every ray is treated as scattered, even
in the specular direction) as coherent scattering, and consequently
both depend upon the surface roughness.  It does not require any
assumptions in order to separate the wave into reflected and scattered
waves, and does not require the small angle approximation so that all
the scattered rays can be traced accurately.

This new study of the century old problem is motivated by our direct
involvement of the evaluation of the X-ray mirror performance aboard
the Chandra X-ray Observatory (CXO) -- the NASA's third great space
observatories, which has been successfully operated since July 23,
1999.  It is the first, and so far the only, X-ray telescope achieving
sub-arcsec angular resolution ($<0.5^{\prime\prime}$ FWHM), which let
us see the X-ray Universe we had never seen before.  Chandra's
spectacular success owes to the genius design and superb manufacture
of its X-ray mirrors.  These mirrors are the largest and the most
precise grazing incidence optics ever built.  At 0.84-m long and 0.6
-- 1.2-m in diameters, the surface area of each mirror ranging from
1.6 to 3.2 square meters.  They were polished to the highest quality
ever achieved for any X-ray mirrors of this size.  The surface
roughness of these mirrors is comparable to or less than the X-ray
wavelengths in the 0.1--10~keV band over most of the mirror surfaces.
However, the mirrors are still not perfect, and consequently there are
still small amount of scattered X-rays.  We need an accurate model of
the X-ray scattering to fully evaluate the Point Spread Function (PSF)
of the CXO in order to correctly understand the scientific data.  In
this paper, we use the Chandra mirror surface roughness data as
examples to illustrate the application of this new scattering method.
This method is also useful for the future generation X-ray
observatories.

\section{RANDOM ROUGH SURFACES}
\label{sec:rrs}

A rough surface is a surface that deviates from the designed or
assumed perfect surface upon which an incident wave achieves perfect
specular reflection without energy loss to any other directions.  In
reality, there is no absolutely perfect surface exist.  Almost all the
surfaces, from the rough ocean to mountainous land, from the deck of
an aircraft carrier to the finest polished mirrors, either natural or
man made surfaces, are rough surfaces.  

Also, no two rough surfaces are identical even they were both formed
in the same well-controlled process.  Not only that, every part of the
same surface is also unique.  We can not predict the exact roughness
on one part of the surface from our knowledge of the other parts of
the same surface.  Its profile is simply ``random''.  These kind of
surfaces are called {\bf random rough surfaces}.  Statistical methods
are required to describe and study them, such as the Power Spectral
Density (PSD, see Section \ref{sec:psd}).  Given the PSD of a
particular surface, we still can not describe the exact roughness of
that surface, but we can describe a series of surfaces with the same
roughness, which reflect and scatter incident waves the same way as
the original surface.

When study wave reflection and scattering, a surface is called
``smooth'' if it specularly reflects the energy of an incident plane
wave into one direction; whereas a surface is called ``rough'' if it
scatters the energy into various directions.  Based on this
definition, the same surface can be called smooth or rough depends on
the wavelength and incident angle.  Rayleigh first studied the sound
wave scattering from rough surfaces in 1877 that led to the
development of the Rayleigh criterion\cite{rayleigh}.  Consider two
parallel adjacent rays, with wavelength $\lambda$, incident on a rough
surface with surface height difference $\sigma$, at a grazing angle
$\alpha$.  Upon reflection in the specular direction, the path
difference between the two rays is
\begin{equation}
\label{eq:rayleigh0}
\Delta L = 2 \sigma sin\alpha
\end{equation}
therefore the phase difference is
\begin{equation}
\label{eq:rayleigh1}
\Delta\varphi = \frac{2\pi}{\lambda}\Delta L = \frac{4\pi\sigma}{\lambda}sin\alpha
\end{equation}

When the surface is perfectly ``smooth'', $\sigma=0$ hence
$\Delta\varphi=0$; two rays are in phase, therefore reflect
specularly.  When $\Delta\varphi=\pi$, two rays cancel each other,
there is no specular reflection; hence the surface is called
``rough''.  In this case, the energy are scattered into other
directions due to the energy conservation.  By arbitrarily choosing
the value halfway between these two extreme cases,
$\Delta\varphi=\pi/2$, we obtain the {\bf Rayleigh criterion} for
smooth surface:
\begin{equation}
\label{eq:rayleigh2}
\frac{4\pi\sigma}{\lambda}sin\alpha < \frac{\pi}{2}~~~~~
\Longrightarrow ~~~~~
\sigma < \frac{\lambda}{8~sin\alpha}
\end{equation}

In reality, there is no clear cut between the so-called ``smooth'' and
``rough''.  A surface satisfying the Rayleigh criterion can only be
considered as ``nearly'' smooth.  Unless $\sigma=0$, there is always a
small amount of energy being scattered into non-specular directions.
As we shall see, the Chandra telescope mirrors do satisfy the Rayleigh
criterion.  But just that small amount of energy scattered away from
the specular direction needs to be taken into account in order to
fully understand its PSF.  In this sense, all the surfaces can be
considered as ``rough'', because there is no real surfaces are perfect
($\sigma=0$).  Table \ref{tab:rrs} gives some examples of ``smooth''
surface based on the Rayleigh criterion.

\begin{table}[h]
\begin{center}
\caption{Examples of ``smooth'' surface based on the Rayleigh criterion}
\label{tab:rrs}
\vspace{0.1in}

\begin{tabular}{c|c|c|c|c} \hline\hline
Surface &  Wave frequency/energy & $\lambda$ & $\alpha$ & $\sigma$ \\ \hline
Airport radar dish & $\sim$3 GHz  & $\sim$10 cm     & $\sim$80$^{\circ}$ & $<$ 1.3 cm \\ 
Satellite TV dish  & $\sim$10 GHz & $\sim$30 mm     & $\sim$60$^{\circ}$ & $<$ 4.3 mm \\ 
Mirror in your bath room & $\sim$545 THz & $\sim$550 nm    &       90$^{\circ}$ & $<$ 69 nm \\
Chandra X-ray Observatory & 0.1 -- 10 keV & 1.24{\rm\AA} -- 124{\rm\AA} & 27.1$^{\prime}$ -- 51.3$^{\prime}$ & $<$ 10{\rm\AA} \\ \hline\hline
\end{tabular}
\end{center}
\end{table}

\section{Power Spectral Density of Rough Surfaces}
\label{sec:psd}

A rough surface is described, statistically, by its surface Power
Spectral Density (PSD) as a function of the surface spatial frequency
$f$.  Consider a 1-dimensional surface with length $L$ and surface
height (i.e.  deviation from a perfectly flat surface): $z=h(x)$.  Its
PSD is defined as:\footnote{The definition $2 W_1$ is conventional,
  where the subscript $_1$ denotes 1-dimensional; the PSD satisfies
  $PSD(-f) = PSD(f)$, and typically positive frequency limits are used
  for most spectral integrals.  The total power, $\sigma^2$, is the
  integral of $2 W_1$ from $f=0$ to $\infty$, i.e.
  $\sigma^2=\int_{0}^{\infty} 2 W_1(f) df$.}
\begin{eqnarray}
\label{eq:psd}
PSD(f) \equiv 2 W_1(f) =
\frac{2}{L}~\left|\int_{-L/2}^{L/2}e^{\imath 2 \pi x f}h(x)dx\right|^2
\end{eqnarray}
The PSD, as it is defined, is the ``spectrum'' of the surface
roughness.  Its value at $f$ is simply the ``power'' at that
frequency.  It is easy to distinguish between periodic and random
rough surfaces from their PSDs.  For periodic rough surfaces, there
are some ``spectral lines'' in their PSDs; whereas there are no lines
exist for a real random rough surface.

Given a PSD function $2 W_1$, the surface roughness amplitude RMS in
the frequency band of $f_1$ -- $f_2$ (both $f_1$ and $f_2$ are
positive) can be calculated as:
\begin{equation}
\label{eq:sigma}
\sigma_{f_1 - f_2}^2  =  \int_{f_1}^{f_2} 2 W_1(f) df
\end{equation}

For a 2-D random rough surface, the two orthogonal dimensions can be
treated separately and each as a 1-D surface. So the above definitions
are still valid.  For example, for the grazing incident rays upon a
2-D random rough surface, one dimension can be chosen as the
intersection of the incident plane and the surface.  The scattered
rays will remain in the incident plane.  This case is called the
in-plane scattering.  The second dimension can be chosen as the
direction on the surface but perpendicular to the incident plane.  In
this case the scattered rays are out of the incident plane.  This case
is called the out-plane, or transverse, scattering.  Since the surface
roughness can be different in these two directions, their PSDs can
also be different.

\section{Chandra X-ray optics}
\label{sec:hrma}

The Chandra X-ray optics -- High Resolution Mirror Assembly (HRMA) --
is an assembly of four nested Wolter Type-I (paraboloid and
hyperboloid) grazing incidence mirrors made of Zerodur and coated with
iridium (Ir)\cite{lvs97,zhao97,zhao98,zhao04}.  The eight mirrors are
named P1,3,4,6 (paraboloid) and H1,3,4,6 (hyperboloid), due to
historical reasons (there were 6 mirror pairs when the HRMA was
designed and two pairs were later removed to reduce the cost).  The
mirror elements were polished by the then Hughes Danbury Optical
Systems, Inc.~(HDOS).  The surface roughness was measured during the
HDOS metrology measurements after the final polishing, but before the
iridium coating\cite{reid95}.  Tests conducted on sample flats before
and after the coating indicate that the coating does not change the
surface roughness.

The instruments used for the measurements were the Circularity and
Inner Diameter Station (CIDS), the Precision Metrology Station (PMS),
and the Micro Phase Measuring Interferometer (MPMI, aka WYKO).  The
CIDS was used to determine the circularity and the inner diameters.
The PMS was used to measure along individual axial meridians.  With
these two instruments, HDOS essentially measured the `hoops' and
`staves' of each mirror barrel, and thus mapped the entire surface.
The micro-roughness was sampled along meridian at different azimuths
using the WYKO instrument at three different magnifications
($\times$1.5, $\times$10 \& $\times$40)\cite{reid95,zhao95}.

These metrology data were Fourier transformed and filtered.  The low
frequency parts of the CIDS and PMS data were used to form mirror
surface deformation (from the designed mirror surface) maps.  The high
frequency parts of the PMS data and the WYKO data were used to
estimate the surface micro-roughness.  Both of them are parts of the
HRMA model we built for the raytrace simulation of the Chandra
performance.

The mirror surface micro-roughness has little variation with azimuth,
but tends to become worse near the mirror ends.  Table \ref{tab:hrma}
\begin{table}[b]
\caption{HRMA Mirror Sections and Their Surface Roughness}
\label{tab:hrma}
\begin{center}
\begin{tabular}{|c|ccccccccccc|c|} \hline \hline
HRMA   &   \multicolumn{11}{c|}{Sections} & Num of  \\ 
Mirror &   \multicolumn{11}{c|}{Surface Roughness Amplitude RMS
$\sigma_{1-1000/{\rm mm}}$ (\AA)} & Sections \\ \hline 
P1 &     & LC & LB & LA & M (88\%) & SA & SB & SC &    &    &    &  7 \\
   &	 &50.3&8.49&4.51& 3.58    &4.91& 5.94&53.9&    &    &    &    \\ \hline
P3 &     &    & LB & LA & M (92\%) & SA & SB &    &    &    &    &  5 \\ 
   &     &    &5.37&5.26&1.96 	   &2.38&4.83&    &    &    &    &    \\ \hline
P4 &     &    & LB & LA & M (93\%) & SA & SB &    &    &    &    &  5 \\ 
   &     &    &6.41&3.15& 2.57     &3.21&6.81&    &    &    &    &    \\ \hline
P6 &     &    & LB & LA & M (94\%) & SA & SB &    &    &    &    &  5 \\ 
   &     &    &37.1&5.23& 3.34     &5.65&20.9&    &    &    &    &    \\ \hline
H1 &  LD & LC & LB & LA & M (88\%) & SA & SB & SC & SD & SE & SF & 11 \\
   & 26.9&5.34&3.64&3.34& 3.32     &3.32&3.32&3.32&3.53&7.30&60.3&    \\ \hline
H3 &     & LC & LB & LA & M (92\%) & SA & SB & SC & SD &    &    &  8 \\
   &     &4.87&2.90&2.23& 2.08     &2.08&2.10&3.95&5.56&    &    &    \\ \hline
H4 &  LD & LC & LB & LA & M (93\%) & SA & SB & SC & SD & SE &    & 10 \\
   & 7.18&3.83&2.61&2.57& 2.36     &2.36&2.74&2.68&4.01&29.4&    &    \\ \hline
H6 &  LD & LC & LB & LA & M (94\%) & SA & SB & SC & SD & SE &    & 10 \\
   & 19.0&4.92&2.51&2.23& 1.95     &1.95&1.95&2.07&2.96&15.9&    &    \\ \hline 
Total &&&&&&&&&&&& 61 \\ \hline\hline
\end{tabular}
\end{center}
\end{table}
shows the surface roughness of the 61 HRMA mirror sections based on
their roughness.  The number underneath each section name is the
surface roughness amplitude RMS, $\sigma_{1-1000/{\rm mm}}$,
calculated according to Eq.~(\ref{eq:sigma}) for $f=1-1000~$mm$^{-1}$.
Each mirror is 838.2~mm in length.  The middle sections (M), which are
the best polished and hence have the lowest PSDs, cover most part of
the mirror surface (the number in parentheses after each M denote the
percentage coverage).  The $\sigma$'s for the M sections are only
1.9--3.6 \AA.  The end sections, where the $\sigma$'s are relatively
higher, cover a very small part of the mirror ($<1\%$), and hence
contribute very little to the mirror performance.

It is seen that all the middle sections are polished to the
satisfaction of the Rayleigh criterion as ``smooth'' surface
(Eq.~\ref{eq:rayleigh2} and Table~\ref{tab:rrs}), therefore they
provide very good reflections in the specular direction for X-rays.

However, since the mirror surfaces are not perfect, there are still
small among of scatterings from the middle sections.  In addition, the
end sections are not ``smooth'' based on the Rayleigh criterion.  So
we need to have a good scattering model in order to understand the PSF
of the telescope.

\begin{figure}
\begin{center}
\includegraphics*[width=5.4in,angle=0]{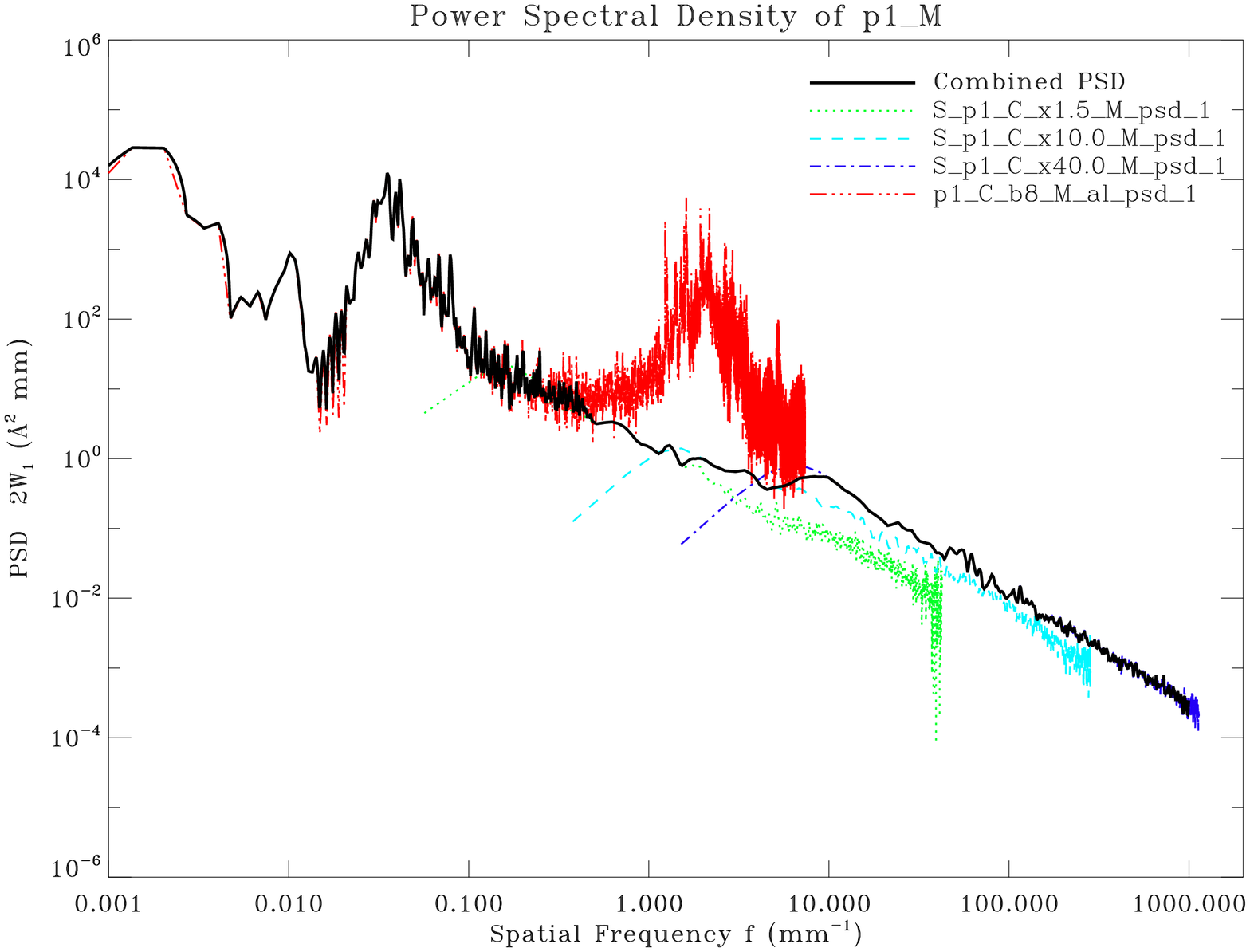}
\caption[psd_p1_m]{ \label{fig:psd_p1_m}
Surface PSD of Chandra mirror P1-M, the middle section of mirror P1.}
\vspace{0.15in}

\includegraphics*[width=5.4in,angle=0]{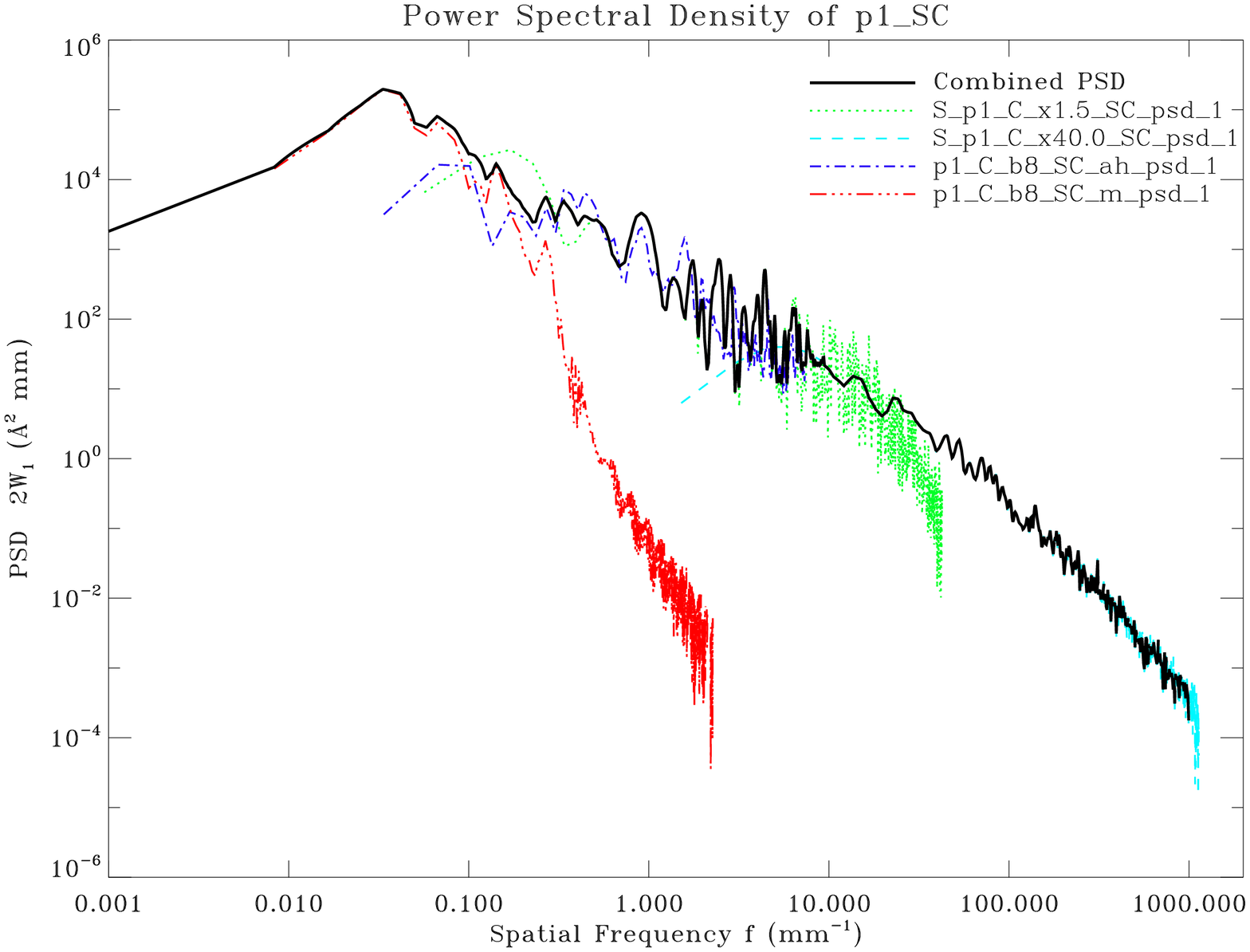}
\caption[psd_p1_sc]{ \label{fig:psd_p1_sc}
Surface PSD of Chandra mirror P1-SC, the small end section of mirror
P1.}
\end{center}
\end{figure}

Figures \ref{fig:psd_p1_m} and \ref{fig:psd_p1_sc} show the PSDs of
the M (middle) and SC (small end) sections of P1.  P1 and H1 are the
first polished mirror pair and are slightly ``rougher'' than other
pairs (see Table \ref{tab:hrma}).  The (colored) dash and dotted lines
show the data from different measurements: the PMS data are in the low
frequency range ($f = 0.001-0.3$~mm$^{-1}$); the WYKO data with 3
magnifications are in the higher frequency range
($f=0.3-1000$~mm$^{-1})$.  The black solid lines are the combined PSDs
from all the measurements.  The SC section obviously is much rougher
than the M section.

\section{ MODEL SURFACES}
\label{sec:surface}

\begin{figure}
\begin{center}
\includegraphics*[trim=0 50 0 40,width=5.4in]{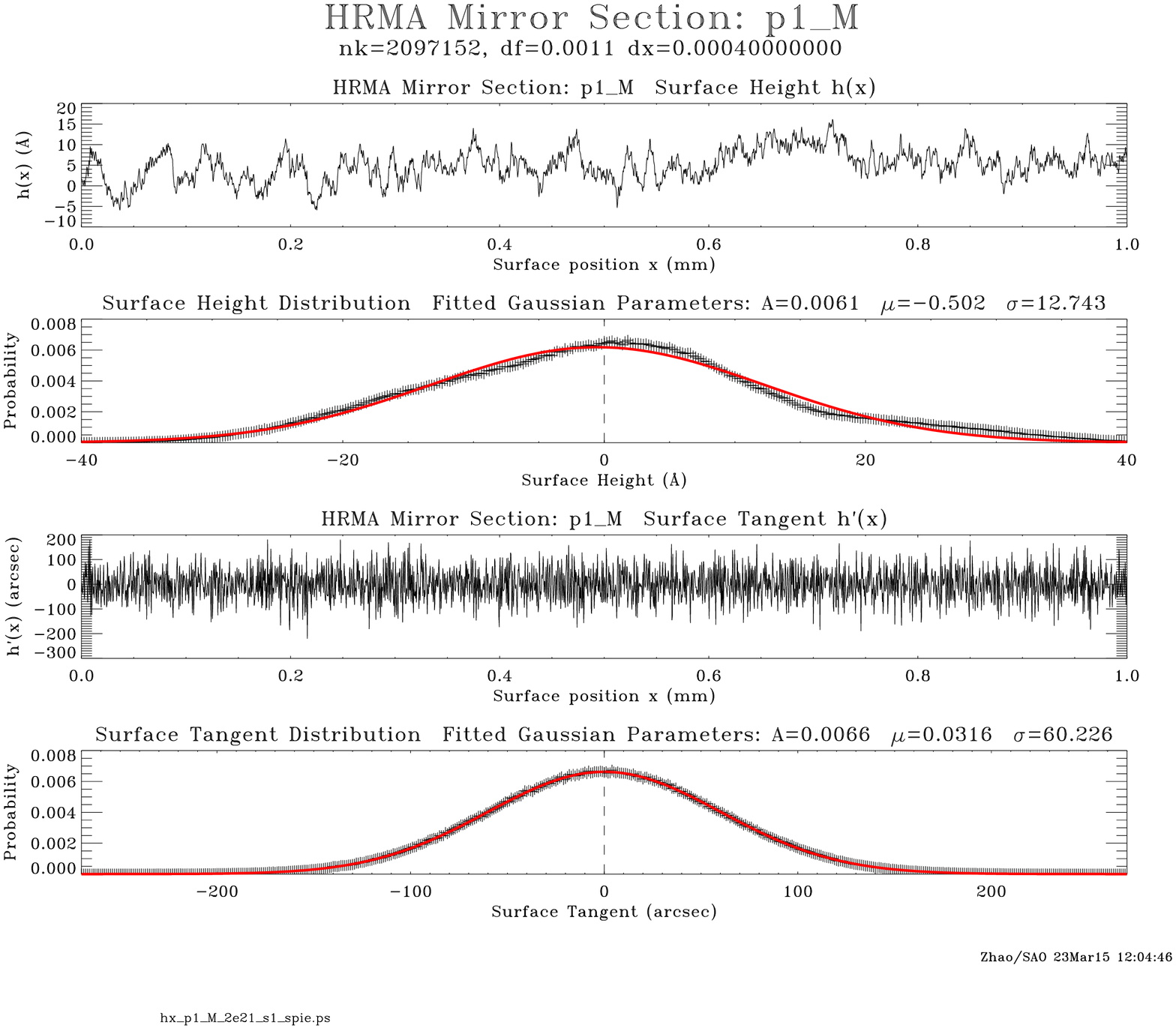}
\vspace{-0.1in}
\caption[surface_p1_m]{\label{fig:surface_p1_m} A model surface of
  mirror section P1-M, which covers 88\% of the P1.  The top panel
  shows the surface height deviation, $h(x)$, for a 1~mm section of
  the model surface; the second panel shows the deviation distribution
  of the entire surface; the third panel shows the surface tangent,
  $h^{\prime}(x)$, for the same section.  The bottom panel shows the
  surface tangent distribution of the entire surface.  Both
  distribution curves match closely with an ideal Gaussian (solid red
  curve).}
\vspace{.1in}

\includegraphics*[trim=0 50 0 40,width=5.4in]{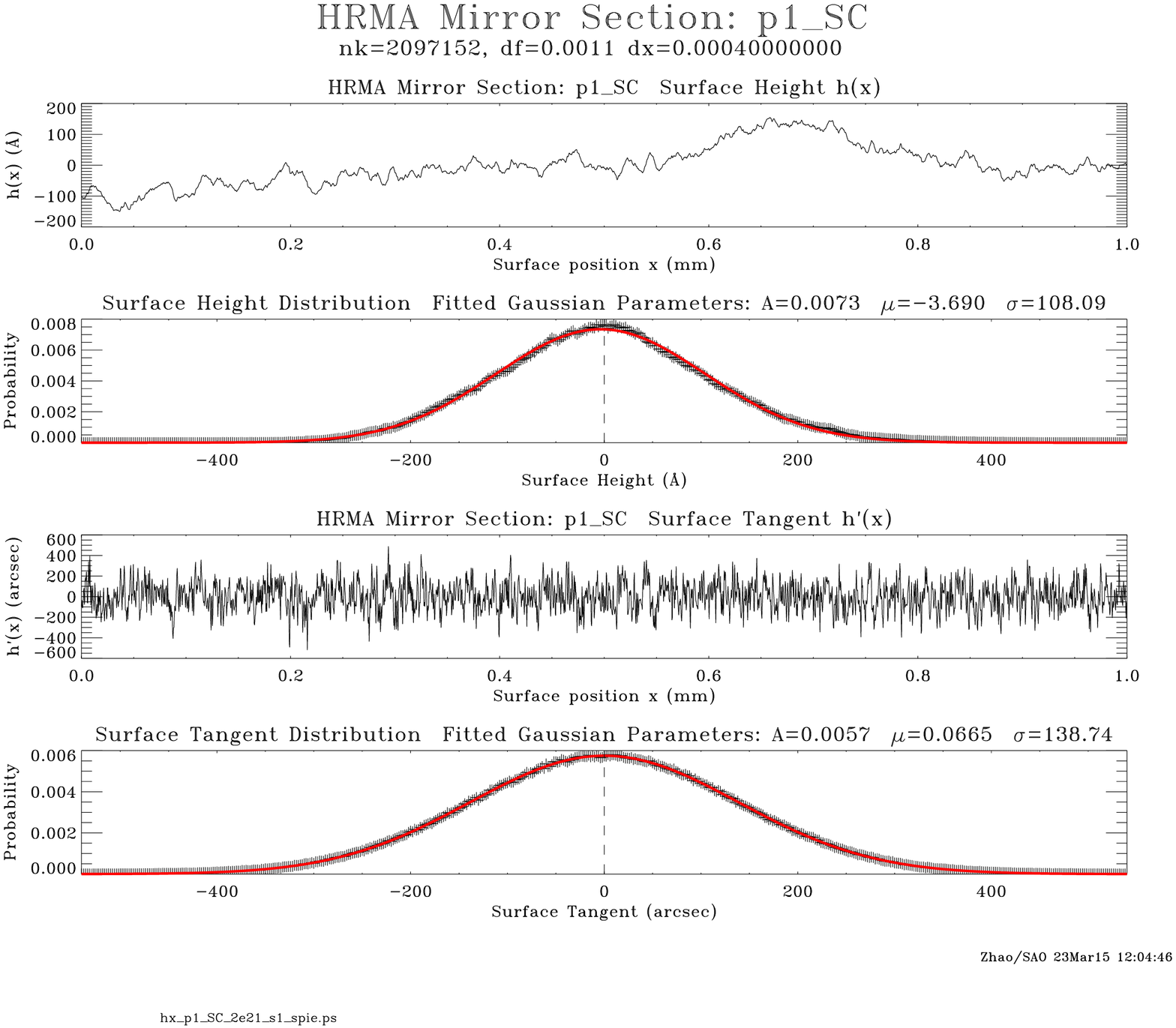}
\vspace{-0.1in}
\caption[surface_p1_sc]{\label{fig:surface_p1_sc} A model surface
  mirror section P1-SC, which is the `worst'' end-section of the P1
  mirror.}
\end{center}
\end{figure}
A random rough surface can be described by its PSD.  Most of the
existing methods calculate the scattering from the surface PSD.
However, our method calculates the scattering directly from the
surface profile.  Therefore, we first need to construct a model
surface that is based on its PSD.  From a random rough surface
profile, we can derive a unique PSD.  But from this unique PSD, we can
not reconstruct the original surface, because the phase information
was lost when deriving the PSD.  However, we can construct any number
of model surfaces with the same roughness as the original one from its
PSD by assigning different random phase factors to the spectral
components.

As mentioned in Section \ref{sec:psd}, for a 2-D surface, we can treat
the two orthogonal dimensions separately and each as a 1-D surface.
To construct a 1-D model surface with length $L$, we need to derive
$N$ consecutive surface heights, $h_i = h(x_i)$, with a fixed interval
$\Delta x$ to cover the surface (i.e. $N\Delta x = L$, where $\Delta
x$ is the spatial resolution of the model surface), and their surface
tangents, $h^{\prime}_i = h^{\prime}(x_i)$.  Appendix A shows that
$h_i$ and $h^{\prime}_i$ can be computed from the surface PSD using
the following Fourier transforms:
\begin{eqnarray}
\label{eq:hi}
h_i &=& \frac{1}{N}\sum^{N/2}_{j=-(N/2-1)}~H_j~e^{-\imath\frac{2\pi
i j}{N}} \\
\label{eq:hid}
h^{\prime}_i &=& \frac{1}{N}\sum^{N/2}_{j=-(N/2-1)}~\left(-\imath 2\pi
f_j~H_j\right)~e^{-\imath\frac{2\pi
i j}{N}}
\end{eqnarray}
where $f_j$ is the surface spatial frequency; $H_j =
N\sqrt{\frac{PSD(f_j)~\Delta f}{2}}~e^{\imath \varphi_j}$, $\Delta f =
1/N\Delta x$, $\varphi_j$ is the assigned random phase factor.  Since
both $h_i$ and $h^{\prime}_i$ are real, this requires $H_{-j} =
H_j^*$, i.e. $PSD(f_{-j})=PSD(f_j)$ and $\varphi_{-j} = -\varphi_j$.

To construct the model surfaces of HRMA, we choose $N=2^{21}$ and
$\Delta x = 0.0004$~mm.  So $L = N\Delta x = 838.86$~mm, and $\Delta f
= 1/N\Delta x = 0.001192$~mm$^{-1}$.  Figures \ref{fig:surface_p1_m}
and \ref{fig:surface_p1_sc} show one set of model surface sections
P1-M and P1-SC, constructed using their PSDs (Fig. \ref{fig:psd_p1_m}
and \ref{fig:psd_p1_sc}) with Eqs.~(\ref{eq:hi}) and (\ref{eq:hid}).

Each HRMA mirror is a 2-D surface.  For any given point on the
surface, its two orthogonal dimensions are the one along the meridian
and the one along the azimuth.  The roughness along the meridian
causes the in-plane scattering.  The roughness along the azimuth
causes the out-plane scattering.  Depending on the polishing method,
the roughness can be different in these two directions.  For the HRMA,
however, the metrology data did not show this difference.  Therefore
we treat these two dimensions as having the same roughness, i.e. the
same PSD is used to construct the model surface profiles in both
directions.
\clearpage
\section{In-plane scattering from model surfaces} 
\label{sec:scatter-in}

In this section, we derive the in-plane scattering formulae of plane
incident waves from a model random rough surface.  Details of the
derivations can be found in Appendices \ref{app:kirchhoff} and
\ref{app:scatter-in}.

We assume the surfaces are sufficiently smooth so that: 1) there is no
shadowing of one part of the surface by another; and 2) there is no
reflection from one part of the surface to another, i.e. there are no
multiple reflections by the same surface.  For an incident plane wave
with grazing angle $\alpha$, the first condition requires that the
absolute values of all the surface tangents, $|h^{\prime}_i|$, are
less than $\alpha$.  The second condition requires $|h^{\prime}_i|$
less than $\alpha/2$ (when $h^{\prime}_i=-\alpha/2$, the reflected
wave is parallel to the surface).  The first condition is
automatically satisfied when the second condition is met.  So the
surface roughness condition for applying this method is:
\begin{eqnarray}
\label{eq:smooth}
|h^{\prime}_i| < \frac{\alpha}{2}
\end{eqnarray}

This condition is easily satisfied for all 61 HRMA sections, as can be
seen by comparing the tangent distributions in Figures
\ref{fig:surface_p1_m} and \ref{fig:surface_p1_sc} with the mean
grazing angles of the four HRMA mirror pairs ($51.26^{\prime},
41.27^{\prime}, 36.43^{\prime}, 27.08^{\prime}$).

The scattering formula is given by Eq.~(\ref{eq:scatteriq}) in
Appendix \ref{app:in-formula} as the discrete Fourier transform of the
field ${\bf E}_i$, on ${\bf x}$-axis, in flat surface $S_0$:
\begin{eqnarray}
\label{eq:scatter-in}
I(\theta_{j+q/p}) &=& \mathcal{A}\,\left(\frac{\Delta
x\,sin(\alpha-\theta_{j+q/p})}{\lambda}\right)^2~
\left|\sum^{N/2}_{i=-(N/2-1)}~\left({\bf E}_i e^{\imath\frac{2\pi i
q/p}{N}}\right)~e^{\imath\frac{2\pi i
j}{N}}\right|^2\hspace{.2in}(q=0,1,\ldots,p-1)
\end{eqnarray}
where the scattering field intensity $I$ is a function of the
scattering angle $\theta_{j+q/p}$, which is the deviation from the
specular reflection direction ($\theta > 0$ is towards the surface,
$\theta < 0$ is aways from the surface); $\lambda$ is the wavelength;
${\bf E}_i$ is the field amplitude, after the reflection, at uniform
grid $x_i$ on ${\bf x}$-axis, where the model surface was constructed.
${\bf E}_i$ is a function of the incident wave, the model surface
height and tangent, and the local reflectivity.  $\mathcal{A}$ is a
normalization factor given by Eq (\ref{eq:factora}).  Again we choose
$N = 2^{21}$ to cover the entire length of the model surface.

\begin{figure}[t]
\includegraphics*[trim=0 0 0 35,width=6.5in]{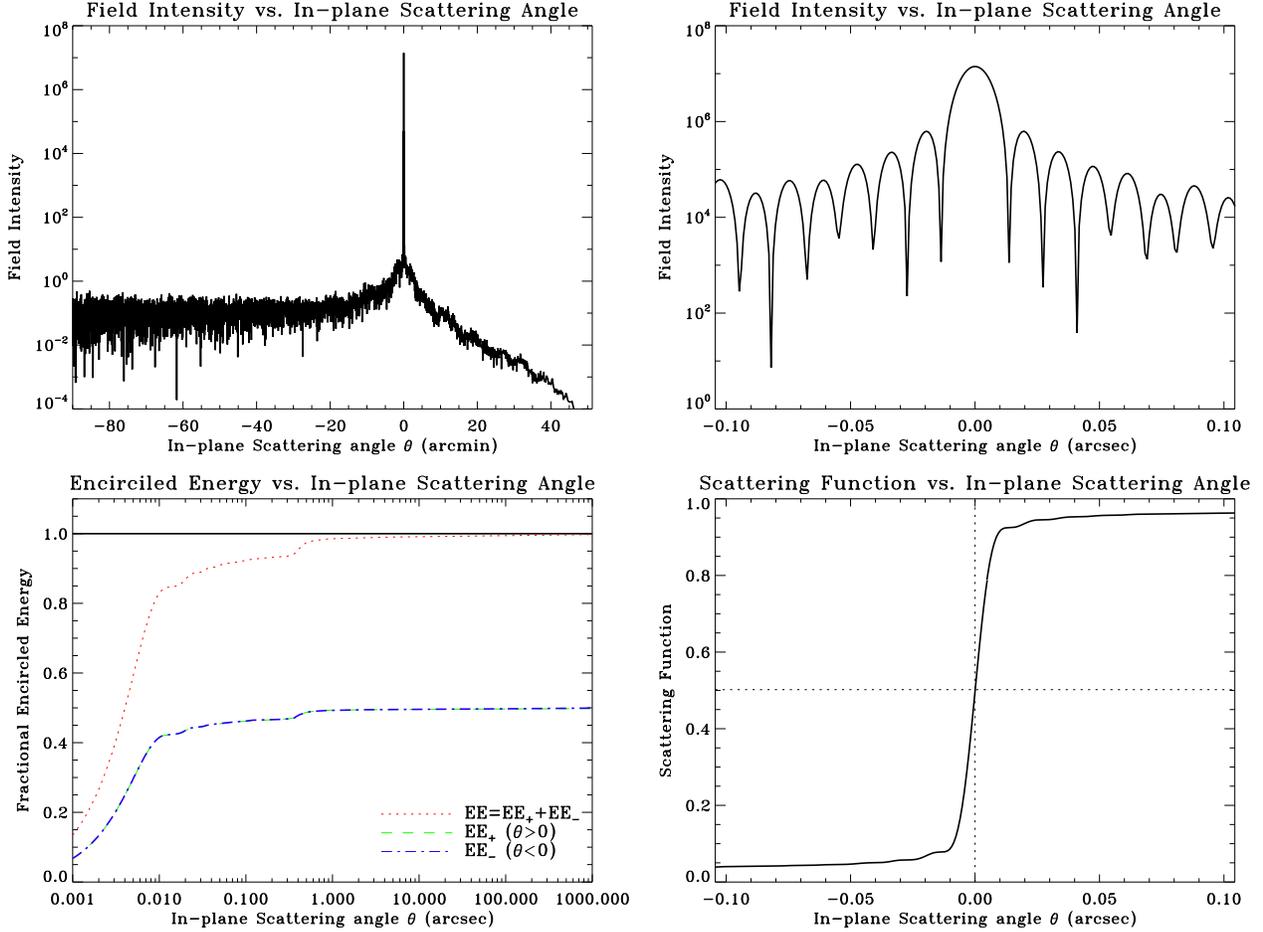}
\caption[scat_p1_m]{ \label{fig:scat_p1_m_in} The in-plane scattering
  of 1.49~keV X-rays at 51.26$^{\prime}$ grazing incident angle from
  the model surface P1-M.  The top-left panel shows the scattering
  field intensity $I$ versus the scattering angle $\theta$.  The very
  sharp peak is at the specular direction $\theta=0$.  The asymmetric
  scattering profile wrt to $\theta=0$ is clearly shown.  The
  top-right panel is the same plot but zoomed into the core of the
  peak; it shows the Fraunhofer diffraction pattern due to the finite
  mirror length.  The bottom-left panel shows the fractional Encircled
  Energy (EE) versus $\theta$, for both sides of the specular
  direction, and also their sum.  The bottom-right panel shows the
  scattering function $\mathcal{S}$ versus $\theta$ in the same range
  as the top-right panel.}
\end{figure}

\begin{figure}[t]
\includegraphics*[trim=0 0 0 35,width=6.5in]{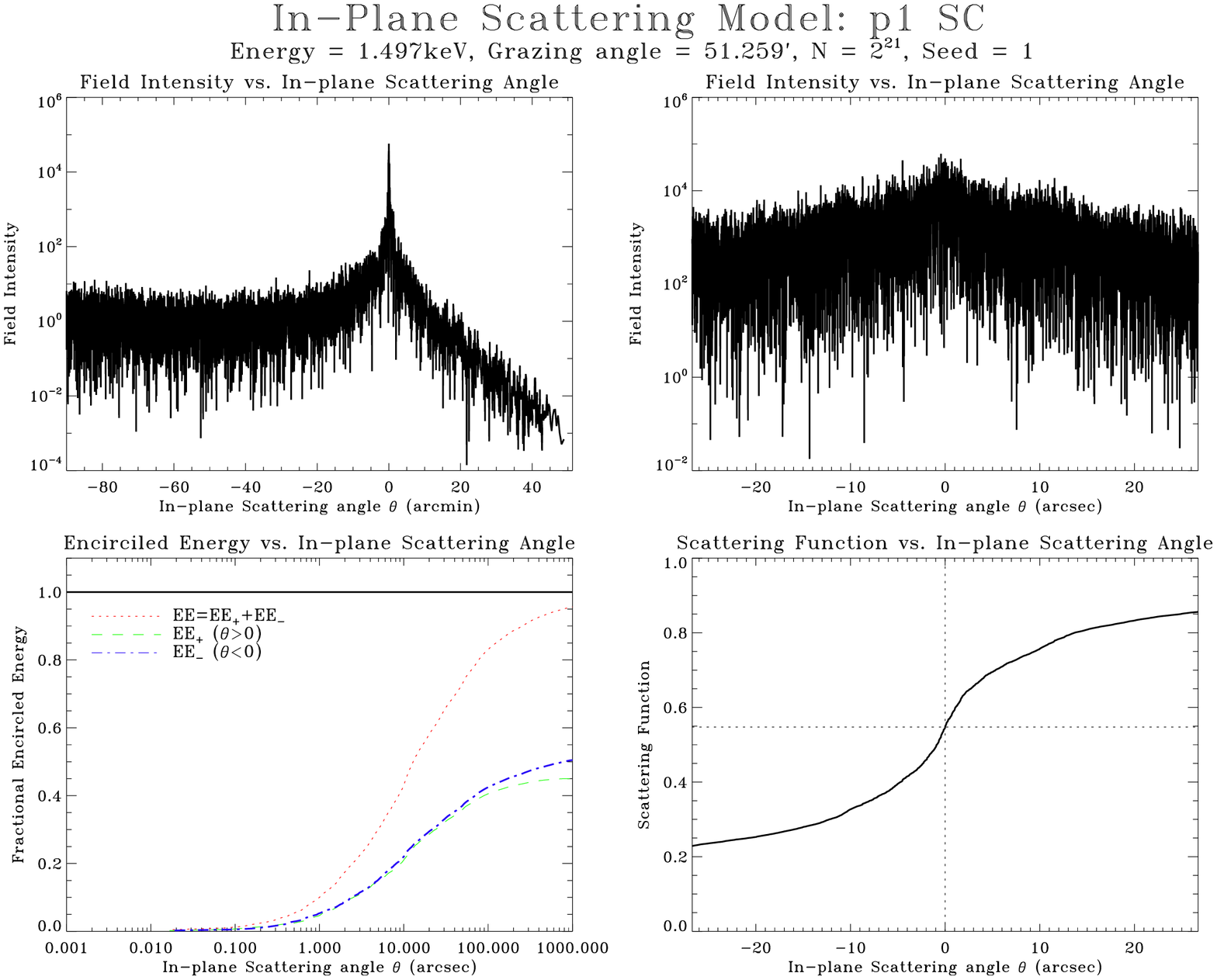}
\caption[scat_p1_sc]{ \label{fig:scat_p1_sc_in} Out-plane scattering
  from model surface P1-SC, with the same description as
  Fig.~\ref{fig:scat_p1_m_in}.  Since P1-SC is much rougher, it has
  much broader scattering profile than P1-M.}
\end{figure}

Figures \ref{fig:scat_p1_m_in} and \ref{fig:scat_p1_sc_in} show the
in-plane scattering results, using Eq (\ref{eq:scatter-in}), for
1.49~keV X-rays incident upon the mirror P1 at its mean grazing angle
(51.26$^{\prime}$).  The top two panels show the scattering intensity
$I$ versus the scattering angle $\theta$.  Positive $\theta$ is
defined as the scattering towards the surface; negative $\theta$ is
the scattering away from the surface.  The sharp peak of specular
reflection (top-left) and the Fraunhofer diffraction pattern
(top-right) are shown as expected.  The bottom two panels show the
fractional Encircled Energies $EE_+, EE_-, EE$ and the scattering
function $\mathcal{S}$, defined as:
\begin{eqnarray}
EE_+(\theta) &\equiv&\frac{1}{\mathcal{E}_s}\int_{0}^{\theta}\,I(\theta_j)\,d\theta_j
~=~ \frac{1}{\mathcal{R}\,\mathcal{E}_i}\int_{0}^{\theta}\,I(\theta_j)\,d\theta_j\\
EE_-(\theta) &\equiv& \frac{1}{\mathcal{E}_s}\int_{-\theta}^{0}\,I(\theta_j)\,d\theta_j
~=~ \frac{1}{\mathcal{R}\,\mathcal{E}_i}\int_{-\theta}^{0}\,I(\theta_j)\,d\theta_j\\
EE(\theta) &\equiv& \frac{1}{\mathcal{E}_s}\int_{-\theta}^{\theta}\,I(\theta_j)\,d\theta_j
~=~ \frac{1}{\mathcal{R}\,\mathcal{E}_i}\int_{-\theta}^{\theta}\,I(\theta_j)\,d\theta_j \\
\label{eq:scatter-in-f}
\mathcal{S}(\theta)&\equiv&\frac{1}{\mathcal{E}_s}\int_{-\pi+\alpha}^{\theta}\,I(\theta_j)\,d\theta_j
~=~ \frac{1}{\mathcal{R}\,\mathcal{E}_i}\int_{-\pi+\alpha}^{\theta}\,I(\theta_j)\,d\theta_j \nonumber \\
\end{eqnarray}
where $\mathcal{E}_i$, $\mathcal{E}_s$ and $\mathcal{R}$ are the total
incident and scattered energy, and the reflectivity of the rough
surface as defined in Appendix \ref{app:in-norm}.  The scattering
function $\mathcal{S}$ is the integral of the scattering intensity $I$
over the angular space $\pi$ above {\bf x}-axis.  It is simply the
probability function (in the domain of [0,1]) of the in-plane
scattering angle $\theta$.

\vspace{0.05in}
\section{Out-plane scattering from model surfaces} 
\label{sec:scatter-out}

When the scattering angle is small comparing to the incident grazing
angle, the transverse scattering angle, i.e.~the out-plane scattering,
is smaller than the in-plane scattering angle by approximately a
factor of the grazing angle.  Therefore traditionally the transverse
scattering was treated by simply multiplying the in-plane scattering
by a factor of the grazing angle $\alpha$ in radians.  This is a good
approximation for small angle scatterings.  However, since our method
is not limited to small angle scatterings, the above approximation is
no longer valid when the scattering angle approaches the grazing
angle.  In this section, we derive the exact equations for the
transverse scattering.  Details of the derivation can be found in
Appendix \ref{app:scatter-out}.

The out-plane scattering formula is given by the discrete Fourier
transform of the field ${\bf E}_i$, on ${\bf y}$-axis, in flat surface
$S_0$, as shown in Eq~(\ref{eq:scatter-outq1}) in
Appendix~\ref{app:out-formula}:
\begin{eqnarray}                                                                
\label{eq:scatter-out}                                                           
J(\eta_{j+q/p})
&=& \mathcal{B}\,\left(\frac{\Delta y\,sin\,\alpha}{\lambda}\right)^2~        
\left|\sum^{N/2}_{i=-(N/2-1)}~\left({\bf E}_i e^{\imath\frac{2\pi i             
q/p}{N}}\right)~e^{\imath\frac{2\pi i j}{N}}\right|^2 \hspace{.3in}(q=0,1,2,\ldots,p-1)
\end{eqnarray}
where the scattering intensity $J$ is a function of the transverse
scattering angle $\eta_{j+q/p}$, which is the deviation from the
specular reflection direction; $\lambda$ is the wavelength; ${\bf
  E}_i$ is the field amplitude, after the reflection, at uniform grid
$y_i$ on ${\bf y}$-axis, where the model surface was constructed.
${\bf E}_i$ is a function of the incident wave, the model surface
height and tangent, and the local reflectivity.  $\mathcal{B}$ is a
normalization factor given by Eq (\ref{eq:factorb}).

\begin{figure}[t]
\includegraphics*[trim=0 0 0 35,width=6.5in]{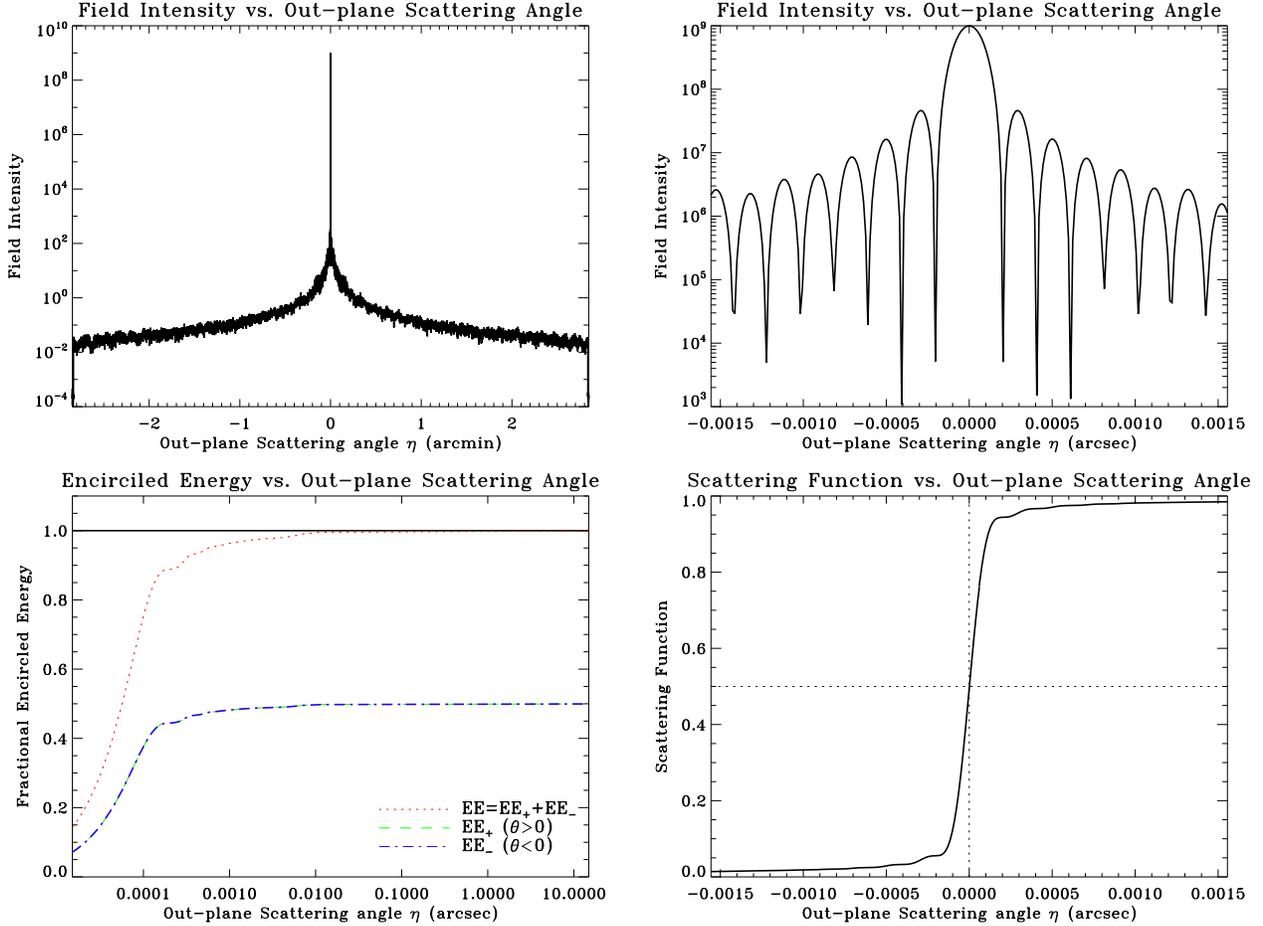}
\caption[scat_p1_m]{ \label{fig:scat_p1_m_out} The out-plane
  scattering of 1.49~keV X-rays at 51.26$^{\prime}$ grazing incident
  angle from the model surface P1-M.  The top-left panel shows the
  transverse scattering field intensity $J$ versus the scattering
  angle $\eta$.  Due to the symmetric nature of the transverse
  scattering, the sign of $\eta$ can be defined in either direction of
  the ${\bf y}$-axis.  The very sharp peak is at the specular
  direction $\eta=0$ (and $\theta=0$).  The transverse scattering
  profile is symmetric wrt to $\eta=0$ as expected.  The top-right
  panel is the same plot but zoomed into the core of the peak; it
  shows the Fraunhofer diffraction pattern due to the finite mirror
  length.  The bottom-left panel shows the fractional Encircled Energy
  (EE) versus $\eta$, for both sides of the specular direction, and
  also their sum.  The bottom-right panel shows the transverse
  scattering function $\mathcal{T}$ versus $\eta$ in the same range as
  the top-right panel.  In all the panels except for the top-left, the
  scale of the X-axis is smaller than that in the same panel of
  Fig.~\ref{fig:scat_p1_m_in} by a factor of the grazing angle
  (51.26$^{\prime}$ = 0.01491 rad).}
\end{figure}

\begin{figure}[t]
\includegraphics*[trim=0 0 0 35,width=6.5in]{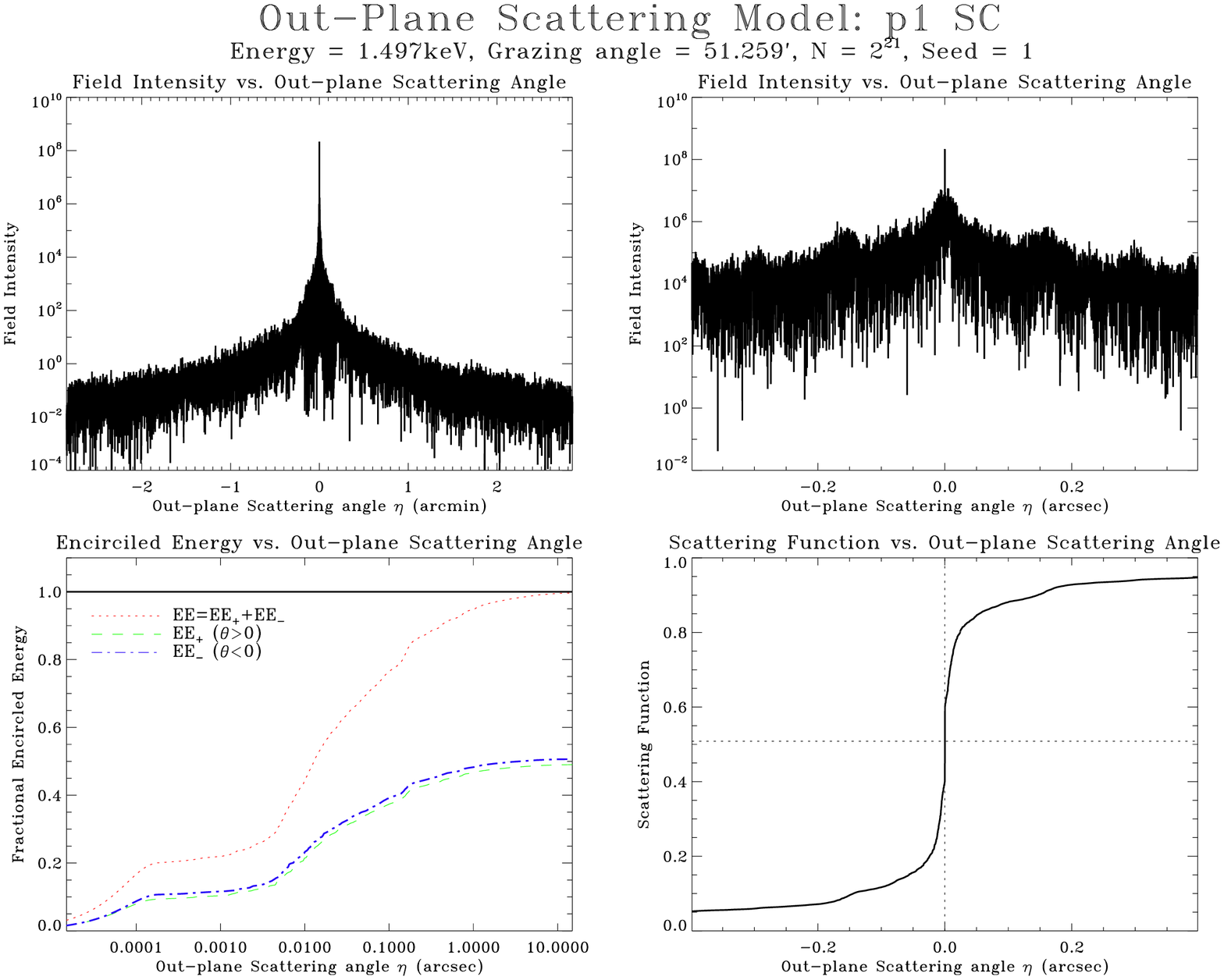}
\caption[scat_p1_sc]{ \label{fig:scat_p1_sc_out} Out-plane scattering from model
  surface P1-SC, with the same description as
  Fig. \ref{fig:scat_p1_m_out}.  It has much broader scattering profile
  than P1-M.}
\end{figure}

Figures \ref{fig:scat_p1_m_out} and \ref{fig:scat_p1_sc_out} show the
out-plane scattering results, using Eq (\ref{eq:scatter-out}), for
1.49~keV X-rays incident upon the mirror P1 at its mean grazing angle
(51.26$^{\prime}$).  The top two panels show the transverse scattering
intensity $J$ versus the scattering angle $\eta$.  The sharp peak
of specular reflection (top-left) and the Fraunhofer diffraction
pattern (top-right) are shown as expected.  The bottom two panels show
the fractional Encircled Energies $EE_+, EE_-, EE$ and the transverse
scattering function $\mathcal{T}$ defined as:
\begin{eqnarray}
EE_+(\eta) &\equiv&\frac{1}{\mathcal{E}_s}\int_{0}^{\eta}\,J(\eta_j)\,d\eta_j
~=~ \frac{1}{\mathcal{R}\,\mathcal{E}_i}\int_{0}^{\eta}\,J(\eta_j)\,d\eta_j\\
EE_-(\eta) &\equiv& \frac{1}{\mathcal{E}_s}\int_{-\eta}^{0}\,J(\eta_j)\,d\eta_j
~=~ \frac{1}{\mathcal{R}\,\mathcal{E}_i}\int_{-\eta}^{0}\,J(\eta_j)\,d\eta_j\\
EE(\eta) &\equiv& \frac{1}{\mathcal{E}_s}\int_{-\eta}^{\eta}\,J(\eta_j)\,d\eta_j
~=~ \frac{1}{\mathcal{R}\,\mathcal{E}_i}\int_{-\eta}^{\eta}\,J(\eta_j)\,d\eta_j\\
\label{eq:scatter-out-f}
\mathcal{T}(\eta)&\equiv&\frac{1}{\mathcal{E}_s}\int_{-\pi/2}^{\eta}\,J(\eta_j)\,d\eta_j
~=~ \frac{1}{\mathcal{R}\,\mathcal{E}_i}\int_{-\pi/2}^{\eta}\,J(\eta_j)\,d\eta_j \nonumber \\
\end{eqnarray}
where $\mathcal{E}_i$, $\mathcal{E}_s$ and $\mathcal{R}$ are the total
incident and scattered energy, and the reflectivity of the rough
surface as defined in Appendix \ref{app:out-norm}.  The scattering
function $\mathcal{T}$ is the integral of the scattering intensity $J$
over the angular space $\pi$ above {\bf y}-axis.  It is simply the
probability function (in the domain of [0,1]) of the out-plane
scattering angle $\eta$.

Now let's compare Fig.~\ref{fig:scat_p1_m_out} of the out-plane
scattering with Fig.~\ref{fig:scat_p1_m_in} of the in-plane
scattering, for the mirror section P1-M.  The top-left panels show the
scattering field intensity versus the scattering angle: the out-plane
scattering is symmetric wrt to the specular direction; while the
in-plane scattering is asymmetric.  This difference is as expected due
to the scattering geometry.  The other three panels show the core of
the intensity, encircled energy and scattering function versus the
scattering angle, with the scale of the X-axis in
Fig.~\ref{fig:scat_p1_m_out} smaller than that of in
Fig.~\ref{fig:scat_p1_m_in} by a factor of the grazing angle
(51.26$^{\prime}$ = 0.01491 rad).  It is seen that these three panels
are very similar to each other in the two figures.  This indicates
that, when the surface is sufficiently ``smooth'', therefore the
scattering angle is small, the out-plane scattering angle is
approximately smaller than the in-plane scattering angle by a factor
of the grazing angle, as shown in Appendix \ref{app:out-small}.

Next let's compare Fig.~\ref{fig:scat_p1_sc_out} with
Fig.~\ref{fig:scat_p1_sc_in}, for the mirror section P1-SC.  The
top-left panels still show the symmetric versus asymmetric profile of
the out-plane and in-plane scattering.  But, because the P1-SC section
is much rougher and therefore scattering angle is larger, the other
three panels look different, even the scale of the X-axis in
Fig.~\ref{fig:scat_p1_sc_out} is still smaller than that of in
Fig.~\ref{fig:scat_p1_sc_in} by a factor of 0.01491.  This indicates
that, when the surface is ``rough'' so that the scattering angles are
large, the small angle approximation treatment for the out-plane
scattering is no longer valid.  An exact solution of of the out-plane
scattering, independent of the in-plane scattering, is required for
the general case.

\section{General solution of scattering from random rough surfaces}
\label{sec:scatter-com}

With the in-plane and out-plane scattering formulae
Eqs~(\ref{eq:scatter-in}) and (\ref{eq:scatter-out}), and their
scattering functions Eqs~(\ref{eq:scatter-in-f}) and
(\ref{eq:scatter-out-f}), the general solution of scattering from
random rough surfaces can be obtained.

The solution can easily be applied in any raytrace or Monte Carlo
simulations.  For a monotonic plane wave with a fixed incident angle
$\alpha$,\footnote{$\alpha$ can be any size and is not limited to
  grazing angle, because this method works for any incident angle.
  For normal incident, $\alpha=90^{\circ}$.  In that case, any two
  orthogonal directions on the surface can be considered as the
  in-plane and out-plane; and the two scattering formulae $I$ and $J$
  are identical.}  if the surface is perfect, the plane wave is
reflected in the specular direction, with the reflecting angle
$\beta = \alpha$.  If the surface is not perfect, then every ray is
treated as a scattered ray, even in the specular direction.

First, two scattering tables are generated using the in-plane and
out-plane scattering functions $\mathcal{S}(\theta)$ and
$\mathcal{T}(\eta)$, for the given photon energy $\varepsilon$
and incident angle $\alpha$.  Then for each ray, two independent
uniform random numbers selected in the domain of [0,1] are used to
find the in-plane and out-plane scattering angles $\theta$ and
$\eta$ from the tabulated $\mathcal{S}(\theta)$ and
$\mathcal{T}(\eta)$ (interpolation is needed in this process).
Finally the reflected ray is deflected from its specular direction by
$\theta$ (in-plane) and $\eta$ (out-plane), respectively.  The
result of this raytrace simulation yields a scattering pattern in a
2-D angular space for the given plane wave.

However, the above process is only good for one fixed energy and
incident angle.  For a real source with an energy spectrum and a range
of incident angles (e.g. photons hit HRMA mirrors at slightly
different angles on the same paraboloid or hyperboloid surface), the
above process needs to be expanded to more general cases.

Each pair of scattering functions $\mathcal{S}(\theta)$ and
$\mathcal{T}(\eta)$ are good only for a given photon energy
$\varepsilon$ and incident angle $\alpha$.  So theoretically, to cover
an energy spectrum with a range of incident angles, many $\mathcal{S}$
and $\mathcal{T}$ are required on a 2-D grid of [$\varepsilon$,
$\alpha$].  Obviously, this requires enormous amount of computation
time and makes the calculation very slow and cumbersome.  However, it
can be shown that, to a very high degree of accuracy, for small
variations of $\varepsilon$ and $\alpha$, $\mathcal{S}$ and
$\mathcal{T}$ only depend on the product of $\varepsilon$ and
$sin\alpha$, instead of depending on them separately.\footnote{This
  can be proven by generating a series of $\mathcal{S}$ and
  $\mathcal{T}$ with different $\varepsilon$ and $\alpha$ but keep the
  product $\varepsilon sin\alpha$ at a fixed value, both $\mathcal{S}$
  and $\mathcal{T}$ almost stay the same.}  This fact greatly reduces
the amount of computations.  Instead of on 2-D, $\mathcal{S}$ and
$\mathcal{T}$ only need to be generated on a 1-D grid of $\varepsilon
sin\alpha$.

Define $\gamma\equiv\varepsilon sin\alpha$. Since
\begin{eqnarray}
\label{eq:lambda}
\lambda = \frac{c}{\nu} = \frac{hc}{\varepsilon}
\end{eqnarray}
where $c$ is the speed of light; $\nu$ is the photon frequency; $h$ is
the Planck constant.  The scattering intensity $I(\theta_j)$ and
$J(\eta_j)$ (Eqs~(\ref{eq:scatter-in}) and (\ref{eq:scatter-out}))
on the 1-D grid of $\gamma$ can be written as:
\begin{eqnarray}
\label{eq:scatter-in1}
I_\gamma(\theta_{j+q/p}) &=& \mathcal{A}\,\left(\frac{\Delta x\, 
    \varepsilon\,sin(\alpha-\theta_{j+q/p})}{hc}\right)^2~
\left|\sum^{N/2}_{i=-(N/2-1)}~\left({\bf E}_{i_{in}} e^{\imath\frac{2\pi i
        q/p}{N}}\right)~e^{\imath\frac{2\pi i
      j}{N}}\right|^2\hspace{.2in}(q=0,1,\ldots,p-1)\hspace{.2in} \\
\label{eq:scatter-out1}                                                           
J_\gamma(\eta_{j+q/p}) &=& \mathcal{B}\,\left(\frac{\Delta y\,
    \varepsilon\, sin\,\alpha}{hc}\right)^2~        
\left|\sum^{N/2}_{i=-(N/2-1)}~\left({\bf E}_{i_{out}} e^{\imath\frac{2\pi i             
        q/p}{N}}\right)~e^{\imath\frac{2\pi i j}{N}}\right|^2 \hspace{0.8in}(q=0,1,\ldots,p-1)\hspace{.2in}
\end{eqnarray}
where ${\bf E}_{i_{in}}$ and ${\bf E}_{i_{out}}$ are the field
amplitudes uniformly distributed on ${\bf x}$ and ${\bf y}$-axis (with
even spacing $\Delta x$ and $\Delta y$) in flat surface $S_0$, for in-
and out-plane scatterings respectively.  They are functions of local
surface height, tangent, reflection coefficient, as well as the photon
energy $\varepsilon$ and incident angle $\alpha$, as explicitly
expressed in Eq~(\ref{eq:ei-in}) in Appendix
\ref{app:in-discrete} and Eq~(\ref{eq:ei-out}) in Appendix
\ref{app:out-discrete}.

The scattering functions $\mathcal{S}(\theta)$ and
$\mathcal{T}(\eta)$ (Eqs~(\ref{eq:scatter-in-f}) and
(\ref{eq:scatter-out-f})) on the 1-D grid of $\gamma$ can be written as:
\begin{eqnarray}                                                                           \label{eq:scatter-in-f1}                                                                   \mathcal{S}_\gamma(\theta) &=& \frac{1}{\mathcal{R}\,\mathcal{E}_i}\int_{-\pi+\alpha}^{\theta}\,I(\theta_j)\,d\theta_j \nonumber\\  
&=& \frac{1}{\mathcal{R}_{in}\,N\varepsilon}\int_{-\pi+\alpha}^{\theta}\,I_\gamma(\theta_j)\,d\theta_j \\
\label{eq:scatter-out-f1}
\mathcal{T}_\gamma(\eta) &=& \frac{1}{\mathcal{R}\,\mathcal{E}_i}\int_{-\pi/2}^{\eta}\,J(\eta_j)\,d\eta_j  \nonumber\\
&=& \frac{1}{\mathcal{R}_{out}\,N\varepsilon}\int_{-\pi/2}^{\eta}\,J_\gamma(\eta_j)\,d\eta_j 
\end{eqnarray}
where $\mathcal{R}_{in}$ and $\mathcal{R}_{out}$ are the in- and
out-plane reflectivities as defined by Eq~(\ref{eq:reflect-in}) in Appendix~\ref{app:in-norm} and
Eq~(\ref{eq:reflect-out}) in Appendix~\ref{app:out-norm}.

The general solution of scattering of plane wave in the photon energy
range [$\varepsilon_0, \varepsilon_1$], ($\varepsilon_0 <
\varepsilon_1$), and incident angle range [$\alpha_0, \alpha_1$],
($\alpha_0 < \alpha_1$), can be obtained in following steps:
\begin{enumerate}
\item Perform Fast Fourier Transform (FFT) to calculate the scattering
  intensities $I_\gamma(\theta_j)$ and $J_\gamma(\eta_j)$, using
  Eqs~(\ref{eq:scatter-in1}) and (\ref{eq:scatter-out1}), on the 1-D
  grid in the domain of $\gamma \in [\varepsilon_0 sin\alpha_0,
  \varepsilon_1 sin\alpha_1]$.  A density of $\Delta\gamma/\gamma
  \approx 2\%$ is sufficient for accurate interpolations.  (e.g. for
  $\frac{\varepsilon_1 sin\alpha_1}{\varepsilon_0 sin\alpha_0} -1
  \approx 0.2$, $\gamma$ assumes 11 values uniformly distributed in
  $[\varepsilon_0 sin\alpha_0, \varepsilon_1 sin\alpha_1]$ is
  sufficient.
\item For each $\gamma$, calculate and tabulate the scattering functions
  $\mathcal{S}_\gamma(\theta)$ and $\mathcal{T}_\gamma(\eta)$,
  using Eqs~(\ref{eq:scatter-in-f1}) and (\ref{eq:scatter-out-f1}).
  The resulting scattering tables can be read as the reverse functions
  of $\theta(\mathcal{S}_\gamma)$ and $\eta(\mathcal{T}_\gamma)$,
  with both $\mathcal{S}_\gamma$ and $\mathcal{T}_\gamma$ assume uniform
  grid of 100,000 points in the domain of [0,1] in order to achieve
  good angular resolutions.
\item For each incident ray with energy $\varepsilon$ and incident
  angle $\alpha$, use two independent uniform random numbers selected
  in the domain of [0,1] and interpolation to find its in-plane and
  out-plane scattering angles $\theta$ and $\eta$ from the
  scattering tables.
\item Finally, the ray is deflected from its specular direction by
  $\theta$ (in-plane) and $\eta$ (out-plane), respectively.
\end{enumerate}
The above process yields a scattering pattern in the 2-D angular space
for the plane wave with a finite range of photon energies and incident
angles.  This completes the general solution of wave scattering from
random rough surfaces.

\section{SUMMARY}
\label{sect:summary}

The exact solution of wave scattering from random rough surfaces are
derived.  This solution provides a new method to solve the long
standing problem of scattering from random rough surfaces in an
accurate and more general way.  This new method treats both the
reflected wave and scattered wave together as coherent scattering,
instead of treating them separately as coherent reflection and
incoherent (diffuse) scattering.  Table \ref{tab:summary} compares
different aspects of the scattering treatment between the traditional
method and the this new method.

\begin{table}[h]
\begin{center}
\caption{Comparing Traditional Method with the New Method of scattering from Random Rough Surfaces}
\label{tab:summary}
\begin{tabular}{c|c|c} \hline\hline
Aspect               & Traditional Method & New Method \\ \hline
Scatter and reflection & treated separately as coherent     & treated together as \\
                     & reflection and diffuse scattering   & coherent scattering \\
Scattered rays       & only some rays are treated as scattered & every ray is treated as scattered \\
Scattering angle     & much smaller than the grazing angle & no restrictions \\ 
In-plane scattering  & symmetric wrt specular direction    & asymmetric wrt specular direction \\
Out-plane scattering & grazing angle times in-plane scattering & solved independently \\ \hline\hline
\end{tabular}
\end{center}
\end{table}

The major advantage of this new method is that it is not limited by
the small angle approximation and gives accurate solutions to in-plane
and out-plane scatterings of any angular size.  This made it generally
applicable in many wave scattering problems.  It is especially useful
for X-ray scattering at grazing angles.  This new method will be very
useful for the future X-ray astronomy missions.

\vspace{0.2in}

\appendix    
\section{Construction of Model Surfaces}
\label{app:surface}

\subsection{Fourier Transform}
\label{app:fourier}
The {\bf Continuous Fourier Transform} equations are\cite{[cf. ]press}:
\begin{eqnarray}
\label{eq:fourier}
H(f) &=& \int^{\infty}_{-\infty}~h(x)~e^{\imath 2\pi x f}~dx
\hspace{1in}(forward)\\
h(x) &=& \int^{\infty}_{-\infty}~H(f)~e^{-\imath 2\pi x f}~df
\hspace{0.9in}(inverse)
\end{eqnarray}
Here if $h$ is a function of position, $x$, in mm, $H$ will be a
function of spatial frequency, $f$, in mm$^{-1}$.

When there are $N$ consecutive sampled values at $x=x_i$ with the
sampling interval $\Delta x$, we make the transform:
\begin{eqnarray}
x &\Rightarrow& x_i \equiv i~\Delta x,~~~~~h(x) \Rightarrow h_i \equiv
h(x_i),~~~~~~i =
-(\frac{N}{2}-1),\ldots,-1,0,1,\ldots,\frac{N}{2}\\
f &\Rightarrow& f_j \equiv j~\Delta f,~~~~H(f) \Rightarrow
H_j \equiv \frac{H(f_j)}{\Delta x},~~~j =
-(\frac{N}{2}-1),\ldots,-1,0,1,\ldots,\frac{N}{2}
\end{eqnarray}
where $\Delta x \Delta f = 1/N$.  We obtain the {\bf Discrete Fourier
Transform} equations:
\begin{eqnarray}
\label{eq:disfor}
H_j &=&\sum^{N/2}_{i=-(N/2-1)}~h_i~e^{\imath\frac{2\pi i
j}{N}}\hspace{1.1in}(forward)\\
\label{eq:disinv}
h_i &=& \frac{1}{N}\sum^{N/2}_{j=-(N/2-1)}~H_j~e^{-\imath\frac{2\pi
i j}{N}}
\hspace{.8in}(inverse)
\end{eqnarray}

\subsection{Surface Height}
\label{app:surhei}
From Eq (\ref{eq:psd}), we obtain:
\begin{eqnarray}
PSD(f) ~=~ \frac{2}{L}~\left|\int_{-L/2}^{L/2}e^{\imath 2 \pi x f}h(x)dx\right|^2
~~~~\Longrightarrow~~~~
\label{eq:psdl2}
\sqrt{\frac{PSD(f)~L}{2}} ~=~ \left|\int_{-L/2}^{L/2}e^{\imath 2 \pi x f}h(x)dx\right|
\end{eqnarray}
Here $PSD(f)$ is a real continuous function of the spatial
frequency $f$.  We first need to convert Eq (\ref{eq:psdl2}) to
a discrete Fourier transform.  Using the equations in \ref{app:fourier}
and relation $L = N\Delta x = 1/\Delta f$, we obtain:
\begin{eqnarray}
|H_j| ~=~ \frac{|H(f_j)|}{\Delta x} ~=~
\frac{1}{\Delta x}\sqrt{\frac{PSD(f_j)~L}{2}} ~=~ N
\sqrt{\frac{PSD(f_j)~\Delta f}{2}} ~=~
\left|\sum^{N/2}_{i=-(N/2-1)}~h_i~e^{\imath\frac{2\pi i
j}{N}}\right| 
\end{eqnarray}
Therefore $H_j$ can be expressed as the forward Fourier
transform of $h_i$ as
\begin{eqnarray}
H_j ~=~ N \sqrt{\frac{PSD(f_j)~\Delta f}{2}} e^{\imath \varphi_j} ~=~
\sum^{N/2}_{i=-(N/2-1)}~h_i~e^{\imath\frac{2\pi i j}{N}}
\end{eqnarray}
Hence the surface height, $h(x_i) ~=~ h_i$, can be expressed as the
inverse Fourier transform of $H_j$
\begin{eqnarray}
h_i ~=~ \frac{1}{N}\sum^{N/2}_{j=-(N/2-1)}~H_j~e^{-\imath\frac{2\pi
i j}{N}} ~=~ \frac{1}{N}\sum^{N/2}_{j=-(N/2-1)}~N
\sqrt{\frac{PSD(f_j)~\Delta f}{2}} e^{\imath \varphi_j}~e^{-\imath\frac{2\pi i j}{N}}
\end{eqnarray}
where $\varphi_j$ is a random phase factor.  A set of surface heights,
$h_i$, can be generated from a set of phase factor $\varphi_j$.
Therefore for a given PSD, we can generate as many sets of surface map
(of the same roughness) as we want by changing the random phase factor
$\varphi_j$.  Because $h_i$, the surface height, has to be real,
this requires $H_{-j} = H_j^*$, i.e. $PSD(f_{-j})=PSD(f_j)$ and
$\varphi_{-j} = -\varphi_j$.

\subsection{Surface Tangent}
\label{app:surtan}
Since
\begin{eqnarray}
\label{eq:surhei1}
h_i ~=~ \frac{1}{N}\sum^{N/2}_{j=-(N/2-1)}~H_j~e^{-\imath\frac{2\pi i j}{N}} ~=~
\frac{1}{N}\sum^{N/2}_{j=-(N/2-1)}~H_j~e^{-\imath 2\pi x_i f_j}
\end{eqnarray}
The surface tangent can be obtained by taking the derivative on both
sides of Eq. (\ref{eq:surhei1}) with respect to  $x_i$:
\begin{eqnarray}
h^{\prime}_i ~=~
\frac{1}{N}\sum^{N/2}_{j=-(N/2-1)}~\left(-\imath 2\pi
f_j~H_j\right)~e^{-\imath 2\pi x_i f_j}
~=~ \frac{1}{N}\sum^{N/2}_{j=-(N/2-1)}~\left(-\imath 2\pi
f_j~H_j\right)~e^{-\imath\frac{2\pi
i j}{N}}
\end{eqnarray}
The surface tangent $h^{\prime}_i$ also has to be real.  This condition is
automatically satisfied because
\begin{eqnarray}
 -\imath 2\pi f_{-j}~H_{-j} ~=~ -\imath 2\pi
(-f_j)~H_j^* ~=~ \imath 2\pi f_j~H_j^* ~=~ (-\imath 2\pi f_j~H_j)^*
\end{eqnarray}

\section{Kirchhoff Solution}
\label{app:kirchhoff}
The wave scattering from random rough surfaces is described by
the Kirchhoff solution\cite{beckmann} and its far-field approximation.

\begin{figure}
\begin{center}
\includegraphics*[trim=90 208 70 150,width=5.5in,angle=0]{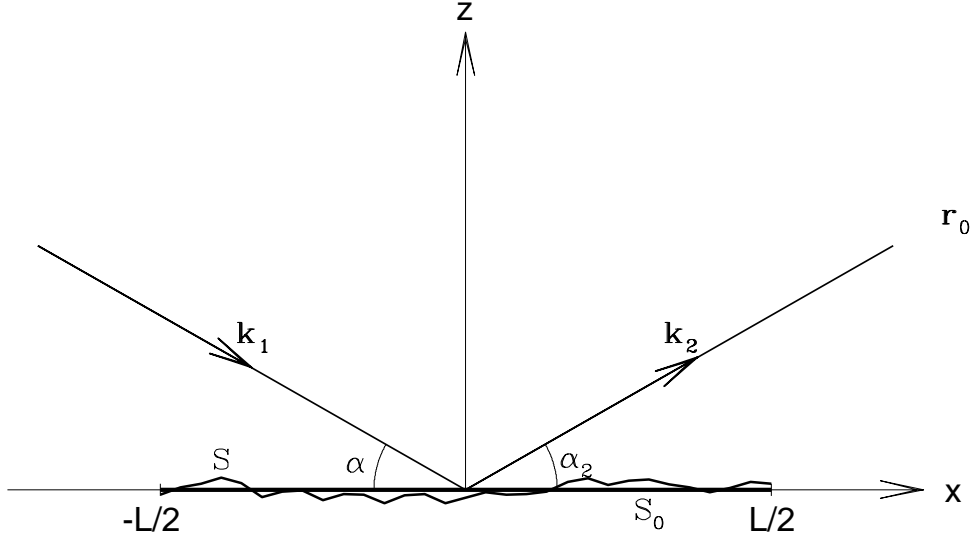}
\caption[surface-in]{\label{fig:surface-in}Wave scattering from a random
rough surface.  A flat surface $S_0$ with $z=0$ lies in the {\bf x-y}
plane ({\bf y}-axis not shown).  A rough surface $S$ has surface
height $z=h(x,y)$, deviates from $S_0$.  The {\bf z} axis is normal to
the {\bf x-y} plane and points up.  Incident and reflecting (or
scattering) wave-vectors are shown as ${\bf k}_1$ and ${\bf k}_2$.  Incident and
reflecting grazing angles with respect to the surface $S_0$ are
$\alpha$ and $\beta$.  ${\bf r}_0$ is the observation point where
the scattering is to be measured.}
\end{center}
\end{figure}
As shown in Figure \ref{fig:surface-in}, define:
\begin{itemize}
\itemsep=-0.02in
\item $S_0$ --- 2-dimensional flat surface at $z=0$.
\item $S$ --- 2-dimensional rough surface, described by its surface
height $z=h(x,y)$.
\item ${\bf E}_1 e^{\imath \mathbf{k_1\cdot r}}~=~{\bf E}_1 e^{\imath
(k_1 x + k_3 z)}$ --- incident plane wave (in the incident plane,
therefore $k_2=0$).
\item ${\bf E}_2 e^{\imath \mathbf{k_2\cdot r}}~=~{\bf E}_2 e^{\imath
(k_x x + k_y y + k_z z)}$ --- reflected or scattered wave from the
rough surface $S$.
\item $\alpha$, $\beta$ --- incident and reflecting grazing
angles with respect to the surface $S_0$.
\end{itemize}
where ${\bf k}_1$ and ${\bf k}_2$ are the wave vectors of the incident
and scattered waves, so ${\bf E}_1\cdot{\bf k}_1=0,~{\bf E}_2\cdot {\bf k}_2=0$, and
\begin{eqnarray}
k ~\equiv~ \frac{2\pi}{\lambda} &=& |{\bf k}_1|~=~\sqrt{k_1^2 +k_3^2}~=~|{\bf k}_2|~=~\sqrt{k_x^2 +k_y^2 +k_z^2} 
\end{eqnarray}

A vector normal to the local surface on $S$ is given by:
\begin{eqnarray}
{\bf n} ~=~ - \nabla (h(x,y)-z)~=~-\frac{\partial h(x,y)}{\partial
x}{\bf\hat x} - \frac{\partial h(x,y)}{\partial y}{\bf\hat y} + {\bf\hat z}
\end{eqnarray}

The field at an observation point ${\bf r}_0$ is given by the
integration of contributions from the field ${\bf E}(s) e^{\imath (k_1
x + k_3 z)}$ on the surface $S$:
\begin{equation}
\label{eq:kirch0}
{\bf E}({\bf r}_0) 
= \frac{1}{\imath \lambda} \int_S \int ds\,
{\bf E}(s) e^{\imath (k_1 x + k_3 z)}\,\frac{e^{\imath k r}}{r^2}\,(\hat{{\bf n}} \cdot {\bf r}) 
= \frac{1}{\imath \lambda} \int\!\!\!\int dx dy \,
{\bf E}(s) e^{\imath (k_1 x + k_3 h(x,y))}\,\frac{e^{\imath k r}}{r^2}\,({\bf n} \cdot {\bf r})
\end{equation}
where $ds$ is an element of surface area; {\bf E}(s) is given by the
incident wave ${\bf E}_1$ multiplied by the suitable reflection
coefficient; the vector ${\bf r}$ goes {\em from} the point of
integration $(x,y,z)$ {\em to} the observation point $(x_0,y_0,z_0)$,
and $r=|{\bf r}|$; $\hat{{\bf n}}$ is a unit vector in the direction
of ${\bf n}$, and $(\hat{{\bf n}} \cdot {\bf r})\, ds=({\bf n} \cdot
{\bf r})\, dx dy$.  Eq. (\ref{eq:kirch0}) is known as the {\bf general
  Kirchhoff solution} for the wave scattering.

Next we derive the far-field approximation of this solution.  When the
reflecting surface is near the origin of the coordinate system and the
observation point is far from the origin, i.e. when $(x\ll x_0,\, y\ll
y_0,\, z\ll z_0)$, we have:
\begin{eqnarray}
{\bf k}_2 &=& k_x {\bf\hat x} + k_y{\bf\hat y}+k_z{\bf\hat z} 
~=~ k\frac{(x_0-x)}{|{\bf r}|} {\bf\hat x} +  k\frac{(y_0-y)}{|{\bf
r}|} {\bf\hat y} +  k\frac{(z_0-z)}{|{\bf r}|} {\bf\hat z}
~\approx~ \frac{k}{r_0}(x_0 {\bf\hat x} +y_0{\bf\hat y}+z_0{\bf\hat z}) \\
{\bf r} &=& (x_0 - x){\bf\hat x} +(y_0 - y){\bf\hat y} + (z_0 -
z){\bf\hat z} ~\approx~ x_0{\bf\hat x} + y_0{\bf\hat y} + z_0{\bf\hat z}
~\approx~ \frac{r_0}{k} \left (k_x{\bf\hat x}+k_y{\bf\hat y}+k_z{\bf\hat z}
\right) \\
r &=& |{\bf r}| ~=~ \sqrt{(x_0 - x)^2 +(y_0 - y)^2 + (z_0 - z)^2}
~\approx~ r_0 - \frac{x_0}{r_0}x - \frac{y_0}{r_0}y- \frac{z_0}{r_0}z
\end{eqnarray}
where $r_0 = |{\bf r}_0|= \sqrt{x_0^2 + y_0^2 + z_0^2}$.  Keep the
first order of $r$ in the phase factor and zeroth order elsewhere:
\begin{eqnarray}
{\bf n} \cdot {\bf r} &\approx& -\frac{r_0}{k}\left[ k_x \frac{\partial
h(x,y)}{\partial x}+k_y \frac{\partial h(x,y)}{\partial y} - k_z
\right] \\
e^{\imath k r} &\approx& e^{\imath k (r_0 - \frac{x_0}{r_0}x -
\frac{y_0}{r_0}y- \frac{z_0}{r_0}z )} ~\approx~ e^{\imath k r_0}\,e^{-\imath
(k_x x + k_y y + k_z h(x,y))}
\end{eqnarray}
Eq. (\ref{eq:kirch0}) becomes:
\begin{eqnarray}
\label{eq:kirchfar1}
{\bf E}({\bf r}_0) &\approx& 
- \frac{1}{\imath \lambda} \int\!\!\!\int dx dy \,
{\bf E}(s) e^{\imath (k_1 x + k_3 z)} \frac{e^{\imath k r_0}}{r_0^2} e^{-\imath (k_x x + k_y y +
k_z z)}\frac{r_0}{k} \left[ k_x \frac{\partial
h(x,y)}{\partial x}+k_y \frac{\partial h(x,y)}{\partial y} - k_z
\right] \\
\label{eq:kirchfar2}
&=&\frac{\imath e^{\imath k r_0}}{2\pi r_0}
\int\!\!\!\int dx dy \,
{\bf E}(s)~e^{\imath (k_1 x + k_3 h(x,y))}~e^{-\imath (k_x x + k_y y
+ k_z h(x,y))}
\left[ k_x \frac{\partial h(x,y)}{\partial x} + k_y \frac{\partial
h(x,y)}{\partial y} - k_z \right]\\
\label{eq:kirchfar3}
&=&\frac{\imath e^{\imath k r_0}}{2\pi r_0}
\int\!\!\!\int dx dy \,
{\bf E}(s) e^{\imath [(k_1-k_x)x - k_y y + (k_3 - k_z) h(x,y)]}
\left[ k_x \frac{\partial h(x,y)}{\partial x} + k_y \frac{\partial
h(x,y)}{\partial y} - k_z \right]
\end{eqnarray}
This is the {\bf far-field approximation} of the Kirchhoff solution
for the wave scattering.

\section{In-plane Scattering formula}
\label{app:scatter-in}

In this section, we derive the in-plane scattering formulae from the
Kirchhoff solution for the constructed model surfaces.

\subsection{Integral on 1-dimensional flat surface $S_0$}
\label{app:in-integral}
\begin{figure}
\begin{center}
\includegraphics*[trim=90 115 70 150,width=5.5in,angle=0]{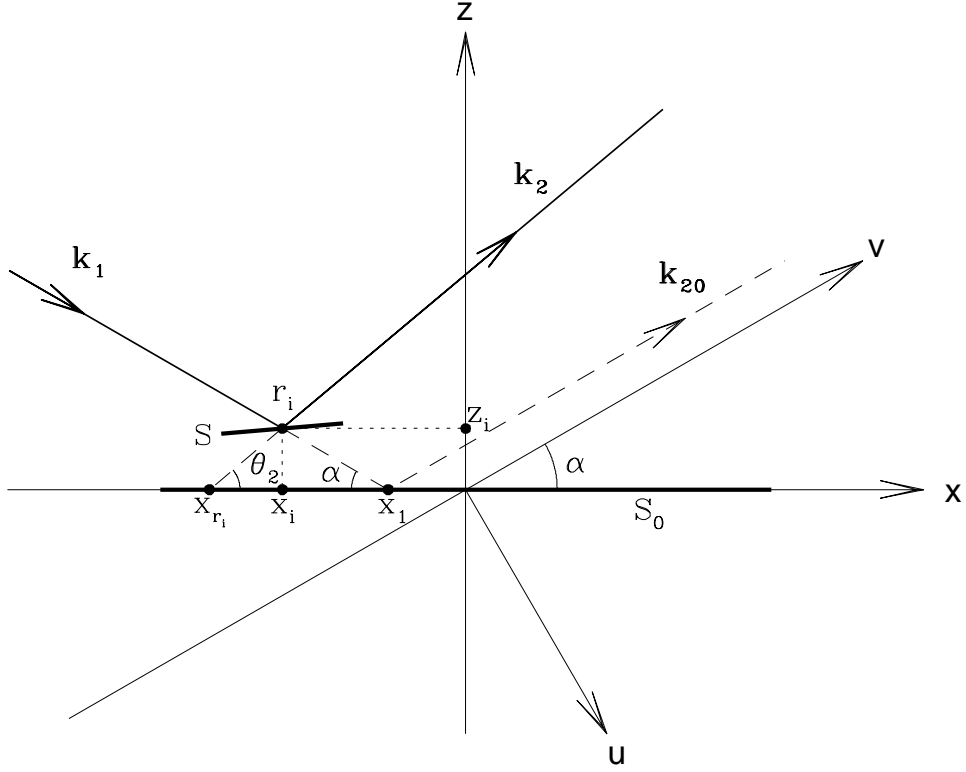}
\caption[scatter]{ \label{fig:scatter-in}The in-plane scattering geometry.  The
flat surface $S_0$ is located on the {\bf x} axis.  The {\bf z} axis is
normal to the surface $S_0$.  The {\bf u-v} axes form a coordinate system
that is rotated clockwise from the {\bf x-z} axes by
$(\frac{\pi}{2}-\alpha)$, so the {\bf v} axis is aligned with the
specular reflection direction.  An incident ray, ${\bf k}_1$, comes in from
the left with a grazing angle $\alpha$; had it struck the surface
$S_0$ at $x_1$, it would have been reflected parallel to the {\bf v}
axis as ${\bf k}_{20}$.  However, it actually strikes the rough surface $S$
at $r_i (x_i,z_i)$, and is reflected at an angle $\beta$ as ${\bf k}_2$.
The intersection of ${\bf k}_2$ with the surface $S_0$ is at $x_{r_i}$.}
\end{center}
\end{figure}

We first isolate the problem by reducing the Kirchhoff solution,
Eq~(\ref{eq:kirchfar1}), to a 1-D integral on ${\bf x}$-axis, in flat
surface $S_0$.  Figure \ref{fig:scatter-in} shows the in-plane
scattering geometry. Consider:
\begin{itemize}
\item For scattering in the incident ({\bf x-z}) plane: $k_y = 0$
\item For 1-D surface in {\bf x} direction, i.e. $h(x,y)$ only depends on $x$:
$h(x,y)=h(x)$
\end{itemize}
Eqs.~(\ref{eq:kirchfar2}) and (\ref{eq:kirchfar3}) become:
\begin{eqnarray}                                                                                         
{\bf E}({\bf r}_0)                                                                                       
\label{eq:in-kirch2}
&\approx&\frac{\imath e^{\imath k r_0}}{2\pi}
\int dx \,{\bf E}(s)~e^{\imath (k_1 x + k_3 h(x))}~e^{-\imath (k_x x+ k_z h(x))}                         
\left[ k_x \frac{dh(x)}{dx} - k_z \right] \\
\label{eq:in-kirch3}
&=&\frac{\imath e^{\imath k r_0}}{2\pi}                                                            
\int dx \,{\bf E}(s)~e^{\imath [(k_1-k_x)x + (k_3 - k_z) h(x)]}                                          
\left[ k_x \frac{dh(x)}{dx} - k_z \right]
\end{eqnarray}
here we have omitted a dimensionless factor $a=Y/r_0$, where $Y$ is
the surface length along the {\bf y}-axis; this factor will be
absorbed later in an overall normalization factor $\mathcal{A}$.

Figure \ref{fig:scatter-in} shows the scattering geometry.  The incident
ray, ${\bf k}_1$, strikes the rough surface $S$ at $r_i (x_i,z_i)$ and is
reflected as ${\bf k}_2$, where $x_i$ is one of the $N$ positions of
the constructed model surface (see Appendix \ref{app:surface}) and
$z_i = h(x_i) = h_i$.  The reflected field at $r_i$ is
\begin{eqnarray}
{\bf E}(s)\,e^{\imath (k_1 x_i + k_3 z_i)} ~=~ {\bf E}(r_i)\,e^{\imath (k_1 x_i + k_3 z_i)} ~=~ {\bf E}(x_i,h_i)\,e^{\imath
(k_1 x_i + k_3 h_i)}
\end{eqnarray}

For the integral (\ref{eq:in-kirch2}), this is equivalent to have a
field at $(x_{r_i},0)$, the intersection of the extension of ${\bf
  k}_2$ and {\bf x} axis, on the surface $S_0$, described by:
\begin{eqnarray}
{\bf E}(x_{r_i},0) ~=~ {\bf E}(r_i)\,e^{\imath (k_1 x_i + k_3 h_i - k h_i/sin\,\beta)}
\end{eqnarray}
where $k h_i/sin\,\beta$ is the phase delay between $(x_i,h_i)$ and
$(x_{r_i},0)$.  Let:
\begin{eqnarray}
{\bf E}(x_{r_i}) ~=~ {\bf E}(x_{r_i},0)\,e^{-\imath k_1 x_{r_i}} ~=~ {\bf E}(r_i)\,e^{\imath
(k_1 x_i + k_3 h_i - k h_i/sin\,\beta - k_1 x_{r_i})} ~=~ {\bf E}(r_i)\,e^{\imath\,\phi_i}
\end{eqnarray}
Substituting the reflected field ${\bf E}(s)\,e^{\imath (k_1 x_i +
  k_3 z_i)}$ at $r_i$ with ${\bf E}(x_{r_i},0)$ at $(x_{r_i},0)$, the
integral (\ref{eq:in-kirch2}) can be written as
\begin{eqnarray}
\label{eq:in-kirch0}
{\bf E}({\bf r}_0)~=~\frac{\imath e^{\imath k r_0}}{2\pi}
\int dx \,{\bf E}(x)~e^{\imath (k_1 x - k_x x -
k_z h(x))}\left[ k_x \frac{dh(x)}{dx} - k_z \right]
\end{eqnarray}

Now the integration boundary has changed from ${\bf E}(s)$ on the
rough surface $S$ to ${\bf E}(x)$ on the flat surface $S_0$, so
$h(x)=0$ and $\frac{dh(x)}{dx} = 0$.  Therefore
Eq. (\ref{eq:in-kirch0}) becomes:
\begin{eqnarray}
\label{eq:in-kirch1}
{\bf E}({\bf r}_0)~=~{\bf E}(k_x,k_z)
~=~\frac{\imath e^{\imath k r_0}}{2\pi}
\int dx \,{\bf E}(x)~e^{\imath (k_1 - k_x) x}~(- k_z)
~=~-\frac{\imath k_z e^{\imath k r_0}}{2\pi}
\int dx \,{\bf E}(x)~e^{\imath (k_1 - k_x) x}
\end{eqnarray}
here the reflected field ${\bf E}(x)$ are calculated at non-uniformly
distributed, discrete points $x=x_{r_i}$.  The position, $x_{r_i}$,
and the phase, $\phi_i$, of the field ${\bf E}(x_{r_i})$ are:
\begin{eqnarray}
\label{eq:ex_pos}
x_{r_i} &=& x_i - \frac{h_i}{tan\,\beta}\\
\label{eq:ex_phase}
\phi_i &=& k_1 x_i + k_3 h_i -\frac{k h_i}{sin\,\beta} - k_1 x_{r_i} 
~=~ k \left(cos\,\alpha x_i - sin\,\alpha h_i -\frac{h_i}{sin\,\beta} -
cos\,\alpha \left(x_i - \frac{h_i}{tan\,\beta}\right)\right)\nonumber \\
&=& - k h_i \left(sin\,\alpha +\frac{1}{sin\,\beta} -
\frac{cos\,\alpha}{tan\,\beta}\right)
~=~-k\,h_i\,\frac{1-cos(\alpha+\beta)}{sin\,\beta}
~=~-2\,k\,h_i\,\frac{sin^2\frac{\alpha+\beta}{2}}{sin\,\beta}
\end{eqnarray}
where $k_3 = -k sin\,\alpha$, because, by definition, the {\bf z}
axis points up; while $k_3$, the $z$ component of the incident ray,
points down.

Thus for the field ${\bf E}(s)$ of each ray ${\bf k}_1$ at $r_i$, we can use
its equivalent field ${\bf E}(x)$ at $x_{r_i}$ to do the integral
($x_{r_i}<x_i$ when $h_i>0$, $x_{r_i}>x_i$ when $h_i<0$).
  
\subsection{Fourier transform with variable $\xi$}
\label{app:in-fourier}
Define a coordinate system {\bf u-v} that is rotated clockwise from
the {\bf x-z} axes by $(\frac{\pi}{2}-\alpha)$, so the {\bf v} axis
is aligned with the specular reflection direction (see Figure
\ref{fig:scatter-in}).  Define the scattering angle, $\theta$, as the
angle of deviation clockwise from the {\bf v} axis, i.e. $\theta =
\alpha-\beta$.  Also define the variable $\xi \equiv
\frac{k_1-k_x}{2 \pi}$.  Therefore:
\begin{eqnarray}
k_1 &=& k~cos\,\alpha,~~~k_x~=~k~cos\,\beta~=~k~cos(\alpha-\theta),
~~~k_z ~=~ k~sin\,\beta ~=~k~sin(\alpha-\theta) \\
\label{eq:xi}
2\pi\xi &=& k_1-k_x ~=~ k~cos\,\alpha - k~cos(\alpha-\theta) ~=~
-2~k~sin(\alpha-\frac{\theta}{2})~sin\,\frac{\theta}{2}\\
\label{eq:theta}
\theta &=&\alpha-cos^{-1}\left(cos\,\alpha-\frac{2\pi\xi}{k}\right)
~=~\alpha-cos^{-1}\left(cos\,\alpha-\xi\lambda \right)
\end{eqnarray}

The scattering equation (\ref{eq:in-kirch1}) becomes:
\begin{eqnarray}
\label{eq:in-scatter0}
{\bf E}({\bf r}_0)&=&{\bf E}(\xi(\theta))
~=~-\frac{\imath k_z e^{\imath k r_0}}{2\pi}\int\,dx\,{\bf E}(x)\, e^{\imath 2\pi\xi x} \\
\label{eq:in-scatter1}
&=&-\frac{\imath e^{\imath k r_0} k~sin(\alpha-\theta)}{2\pi}\int\,dx\,{\bf E}(x)\, e^{\imath 2\pi\xi x}
~=~-\frac{\imath e^{\imath k r_0} sin(\alpha-\theta)}{\lambda}\int\,dx\,{\bf E}(x)\, e^{\imath 2\pi\xi x}
\end{eqnarray}
Thus, the scattering field ${\bf E}(\xi)$ can be obtained from the
Fourier transform integral of the field ${\bf E}(x)$ on the surface
$S_0$.  And it can be expressed as ${\bf E}(\theta)$ via
Eq. (\ref{eq:theta}).

\subsection{Discrete Fourier transform at $x_i$}
\label{app:in-discrete}
In practice, this integral is performed numerically using the Fast
Fourier Transform (FFT) on $N$ uniformly distributed points $x_i$'s
where we constructed the model surface.  Therefore we need to convert
the field ${\bf E}(x_{r_i})$ to the field ${\bf E}(x_i)$.  This can be
simply done by multiplying ${\bf E}(x_{r_i})$ with two factors, $A_i$ and $B_i$:
\begin{eqnarray}
{\bf E}(x_i) ~=~ A_i\, B_i\, {\bf E}(x_{r_i}) ~=~A_i\, B_i\, {\bf E}(x_i
-\frac{h_i}{tan\,\beta}) ~=~A_i\, B_i\,{\bf E}(r_i)\,e^{\imath\phi_i}
\end{eqnarray}
Where the factor $A_i$ is used to adjust the incident plane wave
density due to the different surface height $h_i$'s at the uniform
grid $x_i$'s; it is calculated by intercepting all the incident rays
that strike on the surface $S$ at $(x_i,h_i)$'s with a coordinate,
{\bf w}, that is inside the incident plane and perpendicular to the
direction of incidence.  Let the intercepting points be $w_i$'s on the
coordinate {\bf w}.  Then:
\begin{eqnarray}
A_i~=~ \frac{w_{i+1}-w_{i-1}}{2\, \Delta x\,sin\alpha}
\end{eqnarray}
The factor $B_i$ is used to adjust the outgoing ray density due to the
redistribution of the reflected rays from the non-uniform grid
$x_{r_i}$ to the uniform grid $x_i$.  For example, when the point
$x_{r_i}$ falls between the fixed grid points $x_{i-1}$ and $x_{i}$
($x_{i}-x_{i-1}=\Delta x$), then
\begin{eqnarray}
\frac{x_i-x_{r_i}}{\Delta x}~{\bf E}(x_{r_i})&{\rm is~added~to~field}&{\bf E}(x_{i-1})\\
\frac{x_{r_i}-x_{i-1}}{\Delta x}~{\bf E}(x_{r_i})&{\rm is~added~to~field}&{\bf E}(x_i)
\end{eqnarray}
This process is done for each ray until all the fields are
redistributed to the uniform grid $x_i$.

Having obtained the field ${\bf E}(x_i)$ on uniform grid, $x_i$, we
can rewrite the scattering equation (\ref{eq:in-scatter1}) as the
discrete Fourier transform (see Appendix \ref{app:fourier}).  Let:
\begin{eqnarray}
x &\Rightarrow& x_i \equiv i~\Delta x,~~~~~{\bf E}(x) ~\Rightarrow~ {\bf
E}_i \equiv {\bf E}(x_i),~~~~~~i = -(\frac{N}{2}-1),\ldots,-1,0,1,\ldots,\frac{N}{2}\\
\xi &\Rightarrow& \xi_j \equiv j~\Delta \xi,~~~~~{\bf E}(\xi) ~\Rightarrow~
{\bf E}_j \equiv \frac{{\bf E}(\xi_j)}{\Delta x},~~~~~j =
-(\frac{N}{2}-1),\ldots,-1,0,1,\ldots,\frac{N}{2}
\end{eqnarray}
where $\Delta x \Delta \xi = 1/N$.  The scattering equation
(\ref{eq:in-scatter1}) becomes:
\begin{eqnarray}
{\bf E}_j &\equiv& \frac{{\bf E}(\xi_j)}{\Delta x} 
~=~-\frac{\imath e^{\imath k r_0} sin(\alpha-\theta_j)}{\lambda}
\sum^{N/2}_{i=-(N/2-1)}~{\bf E}_i~e^{\imath\frac{2\pi i j}{N}}
\end{eqnarray}
where
\begin{eqnarray}
\label{eq:ei-in}
{\bf E}_i&=&{\bf E}(x_i)~=~A_i\, B_i\,{\bf E}(r_i)\,e^{\imath\phi_i}
~=~ A_i\, B_i\,{\bf E}_1\,R(\alpha_i)\,e^{\imath\phi_i}
\end{eqnarray}
where ${\bf E}_1$ is the incident plane wave; $R(\alpha_i)$ is the
reflection coefficient of ray $i$ with the local grazing angle,
$\alpha_i$, on the rough surface $S$.  Obviously:
\begin{eqnarray}
\label{betai}
\alpha_i = \alpha+tan^{-1}(h^{\prime}_{xi})
\end{eqnarray}
where $h^{\prime}_{xi}$ ($=dh/dx_i$) is the local surface tangent in the
$\bf x$ direction on the model surface.

The scattering intensity, $I$, is given as a function of the
scattering angle, $\theta$, by:
\begin{eqnarray}
\label{eq:scatteri}
I(\theta_j) ~=~ I(\xi(\theta_j)) ~\equiv~ \mathcal{A}\,{\bf E}(\xi_j)\,{\bf E}^*(\xi_j)
~=~ \mathcal{A}\,\left(\frac{\Delta x\,sin(\alpha-\theta_j)}{\lambda}\right)^2~
\left|\sum^{N/2}_{i=-(N/2-1)}~{\bf E}_i~e^{\imath\frac{2\pi i j}{N}}\right|^2
\end{eqnarray}
where $\mathcal{A}$ is a normalization factor which we will derive in section
\ref{app:in-norm}.

\subsection{Scattering formula -- the Fraunhofer diffraction pattern}
\label{app:in-formula}

With the Eq. (\ref{eq:scatteri}), it seems that we can finally obtain
the profile of scattering from the rough surface.  However, this is
not quite true, because of the discrete Fourier transform.  The main
disadvantage of the discrete Fourier transform is (what else?)
``discrete''.  Its shortcomings are displayed perfectly in this case.
Eq.~(\ref{eq:scatteri}) is correct, but all of the points except the
central peak ($\theta_j=0$) are calculated in the valleys of the
Fraunhofer diffraction pattern at:
\begin{eqnarray}
\theta_j ~=~- \frac{j~\lambda}{N\,\Delta x\,sin\,\alpha}
~=~- \frac{j~\lambda}{L\,sin\,\alpha},~~~~~j=\pm 1, \pm 2, \pm 3, \ldots
\end{eqnarray}
where $L$ is the surface length.  In case of a perfect surface,
Eq.~(\ref{eq:scatteri}) gives $I(\theta_j)=0$ except for one point at
$j=0$, and the correct diffraction pattern from the finite surface
length is not obtained.  To get the diffraction patterns at angles
between $\theta_j$ and $\theta_{j+1}$, we divide
$\theta_{j+1}-\theta_j$ into $p$ equal spaces.  The diffraction
pattern at $\theta_{j+q/p} (q<p)$ can be calculated as:
\begin{eqnarray}
I(\theta_{j+q/p}) 
&=& \mathcal{A}\,\left(\frac{\Delta x\,sin(\alpha-\theta_{j+q/p})}{\lambda}\right)^2~
\left|\sum^{N/2}_{i=-(N/2-1)}~{\bf E}_i~e^{\imath\frac{2\pi i
(j+q/p)}{N}}\right|^2 \hspace{.3in}(q=0,1,2,\ldots,p-1)\\
\label{eq:scatteriq}
&=& \mathcal{A}\,\left(\frac{\Delta x\,sin(\alpha-\theta_{j+q/p})}{\lambda}\right)^2~
\left|\sum^{N/2}_{i=-(N/2-1)}~\left({\bf E}_i e^{\imath\frac{2\pi i
q/p}{N}}\right)~e^{\imath\frac{2\pi i j}{N}}\right|^2 
\end{eqnarray}

So instead of one Fourier transform equation on ${\bf E}_i$, we need
do $p$ Fourier transform equations on ${\bf E}_i\,e^{\imath\frac{2\pi
i q/p}{N}}$.  Usually, $p=16$ is sufficient to calculate very nice
Fraunhofer diffraction patterns.  Eq. (\ref{eq:scatteriq}) is the
final scattering formula.  It maps the field on the surface, $\bf
E(x)$, to the field intensity of scattering, $I(\theta)$.

\subsection{Normalization}
\label{app:in-norm}
Now let's derive the normalization factor $\mathcal{A}$ introduced in
Eq. (\ref{eq:scatteri}).  Let $\varepsilon$ be the energy carried by
each of the $N$ incident rays of the plane wave ${\bf E}_1$.  The
total incident energy, $\mathcal{E}_i$, total reflected energy on the
surface (before scattered away), $\mathcal{E}_r$, and the total
scattered energy (includes all the energies -- reflected and scattered
away from the surface), $\mathcal{E}_s$, are:
\begin{eqnarray}
\mathcal{E}_i &=& N \varepsilon\\
\mathcal{E}_r &=& \sum^{N/2}_{i=-(N/2-1)} \,|{\bf E}_i|^2 ~=~
\varepsilon\sum^{N/2}_{i=-(N/2-1)}\,A_i^2\,B_i^2\,\left|R(\alpha_i)\right|^2\\
\mathcal{E}_s &=& \int d\theta\,I(\theta) ~=~ \mathcal{A}\,\int
d\xi\,|{\bf E}(\xi)|^2
\end{eqnarray}
Define the in-plane reflectivity of the rough surface as:
\begin{eqnarray}
\label{eq:reflect-in}
\mathcal{R} &\equiv& \frac{\mathcal{E}_r}{\mathcal{E}_i} ~=~
\frac{1}{N}\sum^{N/2}_{i=-(N/2-1)}\,A_i^2\,B_i^2\,\left|R(\alpha_i)\right|^2
\end{eqnarray}
With this new method, every reflected ray is considered as the
scattered ray, even it's scattered in the specular direction.  So the
total reflected energy equals to the total scattered energy. Let
$\mathcal{E}_r=\mathcal{E}_s$.  We obtain:
\begin{eqnarray}
\mathcal{A} &=&
\label{eq:factora}
\frac{\varepsilon\sum^{N/2}_{i=-(N/2-1)}\,A_i^2\,B_i^2\,\left|R(\alpha_i)\right|^2}{\int
d\xi\,|{\bf E}(\xi)|^2}
~=~\frac{\varepsilon N \mathcal{R}}{\int d\xi\,|{\bf E}(\xi)|^2} 
~=~\frac{\mathcal{E}_i \mathcal{R}}{\int d\xi\,|{\bf E}(\xi)|^2}
\end{eqnarray}

\section{Out-Plane Scattering formula}
\label{app:scatter-out}

The out-plane (transverse) scattering is due to the surface roughness
in the {\bf y} direction, which is perpendicular to the incident
plane.

\subsection{Integral on 1-dimensional flat surface $S_0$}
\label{app:out-integral}
\begin{figure}
\begin{center}
\includegraphics*[trim=90 115 70 150,width=5.5in,angle=0]{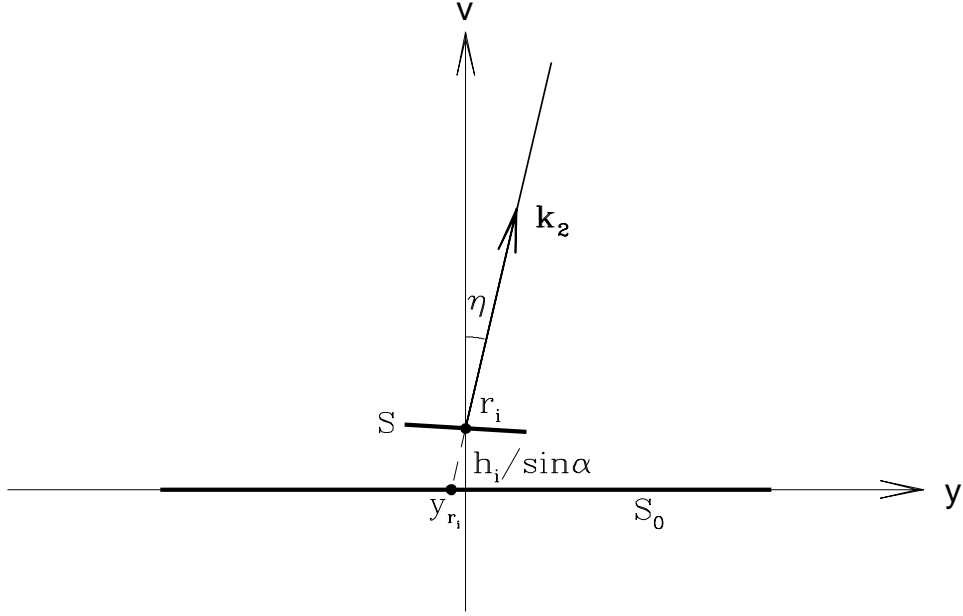}

\caption[scatter_out]{ \label{fig:scatter-out}The out-plane scattering
  geometry.  The 1-D flat surface $S_0$ is located on the {\bf
    y}-axis.  The {\bf v}-axis is the specular direction (see Figure
  \ref{fig:scatter-in}).  The {\bf y-v} plane is perpendicular to the
  incident plane.  The incident ray ${\bf k}_1$, not shown, is behind
  the {\bf v}-axis.  Had it struck the surface $S_0$, it would have
  been reflected in the {\bf v} direction.  However, it actually
  strikes the rough surface $S$ at $r_i (y_i,z_i)$, and is reflected
  at an angle $\eta$ from the {\bf v} direction as ${\bf k}_2$.
  The intersection of the extension of ${\bf k}_2$ with the surface
  $S_0$ is at $y_{r_i}$.}
\end{center}
\end{figure}

Now we isolate the problem by, again, reducing the Kirchhoff solution,
Eq~(\ref{eq:kirchfar1}), to a 1-D integral in flat surface $S_0$, this
time on {\bf y}-axis.  Figure \ref{fig:scatter-out} shows the
out-plane scattering geometry.  Consider:
\begin{itemize}
\itemsep=-0.02in
\item For scattering in the {\bf y-v} plane ({\bf v} is the vector of
  the specular reflection direction): $k_x = k_1$
\item For 1-D surface in {\bf y} direction, i.e. $h(x,y)$ only depends
  on $y$: $h(x,y)=h(y)$
\end{itemize}

Eqs.~(\ref{eq:kirchfar2}) and (\ref{eq:kirchfar3}) become:
\begin{eqnarray}
{\bf E}({\bf r}_0)
\label{eq:out-kirch2}
&\approx&\frac{\imath e^{\imath k r_0}}{2\pi}
\int dy \,{\bf E}(s)~e^{\imath k_3 h(y)}~e^{-\imath (k_y y+ k_z h(y))}
\left[ k_y \frac{dh(y)}{dy} - k_z \right] \\
\label{eq:out-kirch3}
&=&\frac{\imath e^{\imath k r_0}}{2\pi}
\int dy \,{\bf E}(s)~e^{\imath [-k_y y + (k_3 - k_z) h(y)]}
\left[ k_y \frac{\partial h(y)}{\partial y} - k_z \right]
\end{eqnarray}
here we have omitted a dimensionless factor $b=X/r_0$, where $X$ is
the surface length along {\bf x}-axis; this factor will be absorbed
later in an overall normalization factor $\mathcal{B}$.

In Figure \ref{fig:scatter-out}, the incident ray, ${\bf k}_1$,
strikes the rough surface $S$ at $r_i (y_i,z_i)$ and is reflected as
${\bf k}_2$, where $y_i$ is one of the $N$ positions of the
constructed model surface (see Appendix~\ref{app:surface}) and $z_i = h(y_i) =
h_i$.  The reflected field at $r_i$ is
\begin{eqnarray}
{\bf E}(s)\,e^{\imath k_3 z_i} ~=~ {\bf E}(r_i)\,e^{\imath k_3 z_i} ~=~ {\bf E}(y_i,h_i)\,e^{\imath k_3
h_i}
\end{eqnarray}

For the integral (\ref{eq:out-kirch2}), this is equivalent to have a
field at $(y_{r_i},0)$, the intersection of the extension of ${\bf
k}_2$ and {\bf y}-axis, on the surface $S_0$, described by:
\begin{eqnarray}
{\bf E}(y_{r_i},0) ~=~ {\bf E}(r_i)\,e^{\imath
[k_3 h_i - k h_i/(sin\,\alpha\,cos\,\eta)]}
\end{eqnarray}
where $\eta$ is the out-plane scattering angle; $k
h_i/(sin\,\alpha\,cos\,\eta)$ is the phase delay between
$(y_i,h_i)$ and $(y_{r_i},0)$.  Let:
\begin{eqnarray}
{\bf E}(y_{r_i}) ~=~ {\bf E}(y_{r_i},0)\,e^{\imath k h_i/sin\,\alpha} ~=~ {\bf E}(r_i)\,e^{\imath
[k_3 h_i - k h_i/(sin\,\alpha\,cos\,\eta) + k h_i/sin\,\alpha]} ~=~ {\bf E}(r_i)\,e^{\imath\,\psi_i}
\end{eqnarray}
Substituting the reflected field ${\bf E}(s)\,e^{\imath k_3 z_i}$ at
$r_i$ with ${\bf E}(y_{r_i},0)$ at $(y_{r_i},0)$, the integral
(\ref{eq:out-kirch2}) can be written as
\begin{eqnarray}
\label{eq:out-kirch2a}
{\bf E}({\bf r}_0)~=~\frac{\imath e^{\imath k r_0}}{2\pi}
\int dy \,{\bf E}(y)~e^{-\imath [k h(y)/sin\,\alpha + k_y y +k_z h(y)]}\left[ k_y
\frac{dh(y)}{dy} - k_z \right]
\end{eqnarray}

Now the integration boundary has changed from ${\bf E}(s)$ on the
rough surface $S$ to ${\bf E}(y)$ on the flat surface $S_0$, so
$h(y)=0$ and $\frac{dh(y)}{dy} = 0$.  Therefore
Eq (\ref{eq:out-kirch2a}) becomes:
\begin{eqnarray}
\label{eq:out-kirch2b}
{\bf E}({\bf r}_0)~=~{\bf E}(k_y,k_z)
~=~\frac{\imath e^{\imath k r_0}}{2\pi}
\int dy \,{\bf E}(y)~e^{-\imath k_y y}~(- k_z)
\label{eq:out-kirch2c}
~=~-\frac{\imath k_z e^{\imath k r_0}}{2\pi}
\int dy \,{\bf E}(y)~e^{-\imath k_y y}
\end{eqnarray}
here the reflected field ${\bf E}(y)$ are calculated at non-uniformly
distributed, discrete points $y=y_{r_i}$.  The position, $y_{r_i}$,
and the phase, $\psi_i$, of the field ${\bf E}(y_{r_i})$ are:
\begin{eqnarray}
\label{eq:ey_pos}
y_{r_i} &=& y_i - \frac{h_i~ tan\,\eta}{sin\,\alpha}\\
\label{eq:ey_phase}
\psi_i &=& k_3 h_i -\frac{k h_i}{sin\,\alpha\,cos\,\eta} + \frac{k h_i}{sin\,\alpha}
~=~ -k h_i \left(sin\,\alpha+\frac{1}{sin\,\alpha\,cos\,\eta} - \frac{1}{sin\,\alpha} \right) \\
&=& -k h_i~ \frac{sin^2\,\alpha\,cos\,\eta + 1 - cos\,\eta}{sin\,\alpha\,cos\,\eta}
~=~ -k h_i~ \frac{1 - cos^2\,\alpha\,cos\,\eta}{sin\,\alpha\,cos\,\eta}
\end{eqnarray}
where $k_3 = -k sin\,\alpha$, because, by definition, the {\bf
  z}-axis points up; while $k_3$, the $z$ component of the incident
ray, points down.

Thus for the field ${\bf E}(s)$ of each ray ${\bf k}_1$ at $r_i$, we
can use its equivalent field ${\bf E}(y)$ at $y_{r_i}$ to do the
integral.

\subsection{Fourier transform with variable $\zeta$}
\label{app:out-fourier}
Define $\zeta \equiv -k_y/2\pi$.

Since
\begin{eqnarray}
k_y &=& k~sin\,\eta ~=~ -2\pi\zeta
\end{eqnarray}
Therefore
\begin{eqnarray}
\label{eq:out-eta}
\eta &=& sin^{-1}\frac{k_y}{k} ~=~ -sin^{-1}\frac{2\pi\zeta}{k}~=~-sin^{-1}(\zeta\lambda)
\end{eqnarray}

The scattering equation (\ref{eq:out-kirch2b}) becomes:
\begin{eqnarray}
\label{eq:out-scatter0}
{\bf E}({\bf r}_0)~=~{\bf E}(\zeta(\eta))
&=&-\frac{\imath k_z e^{\imath k r_0}}{2\pi} 
\int\,dy\,{\bf E}(y)\, e^{\imath 2\pi\zeta y} \\
\label{eq:out-scatter1}
&=&-\frac{\imath e^{\imath k r_0} k~sin\,\alpha}{2\pi}
\int\,dy\,{\bf E}(y)\, e^{\imath 2\pi\zeta y}
~=~-\frac{\imath e^{\imath k r_0} sin\,\alpha}{\lambda}
\int\,dy\,{\bf E}(y)\, e^{\imath 2\pi\zeta y}
\end{eqnarray}
Thus, the scattering field ${\bf E}(\zeta)$ can be obtained from the
Fourier transform integral of the field ${\bf E}(y)$ on the surface
$S_0$.  And it can be expressed as ${\bf E}(\eta)$ via
Eq (\ref{eq:out-eta}).

\subsection{Discrete Fourier transform at $y_i$}
\label{app:out-discrete}
In practice, this integral is performed numerically using the Fast
Fourier Transform (FFT) on $N$ uniformly distributed points $y_i$'s
where we constructed the model surface.  Therefore we need to convert
the field ${\bf E}(y_{r_i})$ to the field ${\bf E}(y_i)$.  This can be
done by multiplying ${\bf E}(y_{r_i})$ with a factor $C_i$:
\begin{eqnarray}
{\bf E}(y_i) ~=~ C_i\, {\bf E}(y_{r_i}) ~=~C_i\, {\bf E}(y_i
-\frac{h_i~ tan\,\eta}{sin\,\alpha}) ~=~C_i\, {\bf E}(r_i)\,e^{\imath\psi_i}
\end{eqnarray}
where the factor $C_i$ is used to adjust the outgoing ray density due to the
redistribution of the reflected rays from the non-uniform grid
$y_{r_i}$ to the uniform grid $y_i$.  For example, when the point
$y_{r_i}$ falls between the fixed grid points $y_{i-1}$ and $y_{i}$
($y_{i}-y_{i-1}=\Delta y$), then
\begin{eqnarray}
\frac{y_i-y_{r_i}}{\Delta y}~{\bf E}(y_{r_i})&{\rm is~added~to~field}&{\bf E}(y_{i-1})\\
\frac{y_{r_i}-y_{i-1}}{\Delta y}~{\bf E}(y_{r_i})&{\rm is~added~to~field}&{\bf E}(y_i)
\end{eqnarray}
This process is done for each ray until all the fields are
redistributed to the uniform grid $y_i$.

Having obtained the field ${\bf E}(y_i)$ on the uniform grid, $y_i$,
we can rewrite the scattering equation (\ref{eq:out-scatter1}) as the
discrete Fourier transform (see Appendix \ref{app:fourier}).  Let:
\begin{eqnarray}
y &\Rightarrow& y_i \equiv i~\Delta y,~~~~~{\bf E}(y) ~\Rightarrow~ {\bf
E}_i \equiv {\bf E}(y_i),~~~~~~i = -(\frac{N}{2}-1),\ldots,-1,0,1,\ldots,\frac{N}{2}\\
\zeta &\Rightarrow& \zeta_j \equiv j~\Delta \zeta,~~~~~{\bf E}(\zeta) ~\Rightarrow~
{\bf E}_j \equiv \frac{{\bf E}(\zeta_j)}{\Delta y},~~~~~j =
-(\frac{N}{2}-1),\ldots,-1,0,1,\ldots,\frac{N}{2}
\end{eqnarray}
where $\Delta y \Delta \zeta = 1/N$.  The scattering equation
(\ref{eq:out-scatter1}) becomes:
\begin{eqnarray}
{\bf E}_j &\equiv& \frac{{\bf E}(\zeta_j)}{\Delta y} 
~=~-\frac{\imath e^{\imath k r_0} sin\,\alpha}{\lambda}
\sum^{N/2}_{i=-(N/2-1)}~{\bf E}_i~e^{\imath\frac{2\pi i j}{N}}
\end{eqnarray}
where
\begin{eqnarray}
\label{eq:ei-out} 
{\bf E}_i&=&{\bf E}(y_i)~=~C_i\, {\bf E}(r_i)\,e^{\imath\psi_i}~=~ C_i\, {\bf E}_1\,R(\alpha_i)\,e^{\imath\psi_i}
\end{eqnarray}
where ${\bf E}_1$ is the incident plane wave; $R(\alpha_i)$ is the
reflection coefficient of ray $i$ with the local grazing angle,
$\alpha_i$, on the rough surface $S$.  It can be shown:
\begin{eqnarray}
\label{eq:alpha_it}
\alpha_i = sin^{-1}[sin\,\alpha\,cos\,(tan^{-1}(h^{\prime}_{yi}))]
\end{eqnarray}
where $h^{\prime}_{yi}$ ($=dh/dy_i$) is the local surface tangent in the
$\bf y$ direction on the model surface.

The scattering intensity, $J$, can be expressed by the Fourier
transform of field amplitude ${\bf E}_i$, as a function of the
scattering angle, $\eta$:
\begin{eqnarray}
\label{eq:scatter-out0}
J(\eta_j) ~=~ J(\zeta(\eta_j)) ~\equiv~ \mathcal{B}\,{\bf E}(\zeta_j)\,{\bf E}^*(\zeta_j)
~=~ \mathcal{B}\,\left(\frac{\Delta y\,sin\,\alpha}{\lambda}\right)^2~
\left|\sum^{N/2}_{i=-(N/2-1)}~{\bf E}_i~e^{\imath\frac{2\pi i j}{N}}\right|^2
\end{eqnarray}
where $\mathcal{B}$ is a normalization factor which we will derive in section
\ref{app:out-norm}.

\subsection{Out-plane Scattering formula -- the Fraunhofer diffraction
  pattern}
\label{app:out-formula}

For the same reason as described in Appendix \ref{app:in-formula}, the
discrete Fourier transform causes the Eq~(\ref{eq:scatter-out0}) to
compute all the points except the central peak ($\eta_j=0$) in the
valleys of the Fraunhofer diffraction pattern at:
\begin{eqnarray}
\eta_j ~=~- \frac{j~\lambda}{N\,\Delta y}
~=~- \frac{j~\lambda}{L},~~~~~j=\pm 1, \pm 2, \pm 3, \ldots
\end{eqnarray}
where $L$ is the surface length.  In case of a perfect surface,
Eq~(\ref{eq:scatter-out0}) gives $J(\eta_j)=0$ except for one point at
$j=0$, and the correct diffraction pattern from the finite surface
length is not obtained.  To get the diffraction patterns at angles
between $\eta_j$ and $\eta_{j+1}$, we divide
$\eta_{j+1}-\eta_j$ into $p$ equal spaces.  The diffraction
pattern at $\eta_{j+q/p} (q<p)$ can be calculated as:
\begin{eqnarray}
\label{eq:scatter-outq0}
J(\eta_{j+q/p}) 
&=& \mathcal{B}\,\left(\frac{\Delta y\,sin\,\alpha}{\lambda}\right)^2~
\left|\sum^{N/2}_{i=-(N/2-1)}~{\bf E}_i~e^{\imath\frac{2\pi i
(j+q/p)}{N}}\right|^2 \hspace{.3in}(q=0,1,2,\ldots,p-1)\\
\label{eq:scatter-outq1}
&=& \mathcal{B}\,\left(\frac{\Delta y\,sin\,\alpha}{\lambda}\right)^2~
\left|\sum^{N/2}_{i=-(N/2-1)}~\left({\bf E}_i e^{\imath\frac{2\pi i
q/p}{N}}\right)~e^{\imath\frac{2\pi i j}{N}}\right|^2 
\end{eqnarray}
So instead of one Fourier transform on ${\bf E}_i$, we need to
do $p$ Fourier transforms  on ${\bf E}_i\,e^{\imath\frac{2\pi
    i q/p}{N}}$.  Usually, $p=16$ is sufficient to calculate very nice
Fraunhofer diffraction patterns.  Eq~(\ref{eq:scatter-outq1}) is the
final transverse scattering formula.  It maps the field from the
surface, $\bf E(y)$, to the field intensity of scattering,
$J(\eta)$.

\subsection{Normalization}
\label{app:out-norm}
Now let's derive the normalization factor $\mathcal{B}$ introduced in
Eq~(\ref{eq:scatter-out0}).  Let $\varepsilon$ be the energy carried by
each of the $N$ incident rays of the plane wave ${\bf E}_1$.  The
total incident energy, $\mathcal{E}_i$, total reflected energy on the
surface (before scattered away), $\mathcal{E}_r$, and the total
scattered energy (includes all the energies -- reflected and scattered
away from the surface), $\mathcal{E}_s$, are:
\begin{eqnarray}
\mathcal{E}_i &=& N \varepsilon\\
\mathcal{E}_r &=& \sum^{N/2}_{i=-(N/2-1)} \,|{\bf E}_i|^2 ~=~
\varepsilon\sum^{N/2}_{i=-(N/2-1)}\,C_i^2\,\left|R(\alpha_i)\right|^2\\
\mathcal{E}_s &=& \int d\eta\,J(\eta) ~=~ \mathcal{B}\,\int
d\zeta\,|{\bf E}(\zeta)|^2
\end{eqnarray}
Define the reflectivity of the rough surface as:
\begin{eqnarray}
\label{eq:reflect-out}
\mathcal{R} &\equiv& \frac{\mathcal{E}_r}{\mathcal{E}_i} ~=~
\frac{1}{N}\sum^{N/2}_{i=-(N/2-1)}\,C_i^2\,\left|R(\alpha_i)\right|^2
\end{eqnarray}
With this new method, every reflected ray is considered as the
scattered ray, even it's scattered in the specular direction.  So the
total reflected energy equals to the total scattered energy. Let
$\mathcal{E}_r=\mathcal{E}_s$.  We obtain:
\begin{eqnarray}
\mathcal{B} &=&
\label{eq:factorb}
\frac{\varepsilon\sum^{N/2}_{i=-(N/2-1)}\,C_i^2\,\left|R(\alpha_i)\right|^2}{\int
d\zeta\,|{\bf E}(\zeta)|^2}
~=~\frac{\varepsilon N \mathcal{R}}{\int d\zeta\,|{\bf E}(\zeta)|^2} 
~=~\frac{\mathcal{E}_i \mathcal{R}}{\int d\zeta\,|{\bf E}(\zeta)|^2}
\end{eqnarray}

\subsection{Small Angle Approximation}
\label{app:out-small}
Eq (\ref{eq:out-scatter0}) is the exact solution of the out-plane
scattering.  Now it's easy to prove its small angle approximation.
Comparing Eq~(\ref{eq:out-scatter0}) with Eq~(\ref{eq:in-scatter0}) of
the in-plane scattering equation, we found that the only difference is
$\zeta$ instead of $\xi$ in the Fourier transformation.  Consider:
\begin{eqnarray}             
\label{eq:out-nuxi}
\zeta \sim \xi & \Longrightarrow &  -k_y \sim k_1 - k_x  
\end{eqnarray}
But (see Appendix \ref{app:in-fourier})
\begin{eqnarray}                    
k_1-k_x &=& k~cos\,\alpha - k~cos(\alpha-\theta) ~=~
-2~k~sin(\alpha-\frac{\theta}{2})~sin\,\frac{\theta}{2} \\
-k_y &=& -k~sin\,\eta 
\end{eqnarray}
For $\theta \ll \alpha$ and $\eta \ll \alpha$:
\begin{eqnarray}
k_1-k_x ~\approx~ -k~\alpha\theta\\
-k_y ~\approx~ -k~\eta
\end{eqnarray}
Therefore, for scattering angles much smaller than the grazing angle:
\begin{eqnarray}
\eta \sim \alpha\theta
\end{eqnarray}
This proves: {\it When the scattering angles are much smaller than the
  grazing angle, the out-plane scattering angle ($\eta$) is smaller
  than the in-plane scattering angle ($\theta$) by a factor of the
  grazing angle ($\alpha$).}

When $\theta \not\ll \alpha$ and/or $\eta \not\ll \alpha$, a general
solution of the out-plane scattering, Eq (\ref{eq:scatter-outq1}), is
needed.


\vspace{0.2in}

\acknowledgments     

I would like to thank the late telescope scientist of Chandra X-ray
Observatory -- Dr. Leon P. van Speybroeck, who introduced me to this
scattering problem.  His genius made NASA's Chandra X-ray Observatory
a spectacular success.  This work is supported in part by the Chandra
X-ray Center.

\vspace{0.2in}

\bibliography{}   

\begin{thebibliography}{1}

\bibitem{zhao03} 
P.~Zhao and L.~P. van Speybroeck, ``A new method to model X-ray
    scattering from random rough surfaces,'' in {\em X-Ray and
      Gamma-Ray Telescopes and Instruments for Astronomy}, Truemper
    and Tananbaum, eds., {\em Proc. SPIE} {\bf 4815}, p.~124, 2003.

\bibitem{zhao15} Ping~Zhao , ``Transverse X-ray scattering on random
    rough surfaces,'' in {\em EUV and X-Ray Optics: Synergy Between
      Laboratory and Spave IV}, Hudec and Pina, eds., {\em
      Proc. SPIE} {\bf 9510}, 951009, 2015.

\bibitem{rayleigh}
Lord~Rayleigh, {\em The Theory of Sound}, Macmillan, New York, 1877.

\bibitem{beckmann}
P.~Beckmann and A.~Spizzichino, {\em The Scattering of Electromagnetic Waves
  from Rough Surfaces}, Pergamon Press, Oxford, England, 1963.

\bibitem{ogilvy}
See ref. in J.~A. Ogilvy, {\em Theory of Wave Scattering from Random Rough Surfaces}, IOP
  Publishing, Bristol, UK, 1991.

\bibitem{simonsen} I.~Simonsen, ``Optics of Surface Disordered
    Systems'', {\em Eur. Phys. J. Special Topics} 181, 1-103, 2010

\bibitem{lvs97}
L.~P. van Speybroeck, D.~Jerius, R.~J. Edgar, T.~J. Gaetz, and P.~Zhao,
  ``Performance expectation versus reality,'' in {\em Grazing Incidence and
  Multilayer X-ray Optical Systems},  Hoover and Walker, eds., {\em Proc. SPIE}
  {\bf 3113}, p.~89, 1997.

\bibitem{zhao97}
P.~Zhao, L.~M. Cohen, and L.~P. van Speybroeck, ``{AXAF HRMA} mirror ring focus
  measurements,'' in {\em Grazing Incidence and Multilayer X-ray Optical
  Systems},  Hoover and Walker, eds., {\em Proc. SPIE} {\bf 3113}, p.~106,
  1997.

\bibitem{zhao98}
P.~Zhao {\em et~al.}, ``{AXAF} mirror effective area calibration using the
  {C}-continuum source and solid state detectors,'' in {\em X-Ray Optics,
  Instruments, and Missions},  Hoover and Walker, eds., {\em Proc. SPIE} {\bf
  3444}, p.~234, 1998.

\bibitem{zhao04} P.~Zhao {\em et~al.}, ``Chandra X-ray Observatory
  Mirror Effective Area,'' in {\em X-Ray and Gamma-Ray Instrumentation
    for Astronomy XIII}, Flanagan and Siegmund, eds., {\em Proc. SPIE}
  {\bf 5165}, p.~482, 2004.

\bibitem{reid95}
P.~B. Reid, ``Fabrication and predicted performance of the {A}dvanced {X}-ray
  {A}strophysics {F}acility mirror ensemble,'' in {\em X-ray and Extreme
  Ultraviolet Optics},  Hoover and Walker, eds., {\em Proc. SPIE} {\bf 2515},
  p.~361, 1995.

\bibitem{zhao95}
P.~Zhao and L.~P. van Speybroeck, ``{AXAF VETA-I} mirror x-ray test results
  cross check with the {HDOS} metrology data,'' in {\em X-ray and Extreme
  Ultraviolet Optics},  Hoover and Walker, eds., {\em Proc. SPIE} {\bf 2515},
  p.~391, 1995.

\bibitem{press}
cf. W.~H.~Press, S.~A. Teukolsky, W.~T. Vetterling, and B.~P. Flannery, {\em
  Numerical Recipes in C: The Art of Scientific Computing}, Cambridge
  University Press, Cambridge, 1993.

\end{thebibliography}
\bibliographystyle{spiebib}   

\end{document}